\documentclass[aps,pra,twocolumn,superscriptaddress]{revtex4}
\usepackage{graphics}
\usepackage{amsmath}
\usepackage{graphicx}
\usepackage{amsfonts}
\usepackage{amssymb}
\usepackage{float}
\usepackage{longtable}
\usepackage{epsfig}
\usepackage{latexsym}
\usepackage{theorem}
\usepackage{bbm}
\usepackage{bm}
\usepackage{psfrag}
\newtheorem{theorem}{Theorem}
\newtheorem{lemma}[theorem]{Lemma}

\newtheorem{remark}[theorem]{Remark}
\newtheorem{criterion}[theorem]{Criterion}
\begin{document}
\title{Theory of channel simulation and bounds for private communication}
\author{Stefano Pirandola}
\affiliation{Computer Science and York Centre for Quantum Technologies, University of York,
York YO10 5GH, UK}
\author{Samuel L. Braunstein}
\affiliation{Computer Science and York Centre for Quantum Technologies, University of York,
York YO10 5GH, UK}
\author{Riccardo Laurenza}
\affiliation{Computer Science and York Centre for Quantum Technologies, University of York,
York YO10 5GH, UK}
\author{Carlo Ottaviani}
\affiliation{Computer Science and York Centre for Quantum Technologies, University of York,
York YO10 5GH, UK}
\author{Thomas P. W. Cope}
\affiliation{Computer Science and York Centre for Quantum Technologies, University of York,
York YO10 5GH, UK}
\author{Gaetana Spedalieri}
\affiliation{Computer Science and York Centre for Quantum Technologies, University of York,
York YO10 5GH, UK}
\affiliation{Research Laboratory of Electronics, Massachusetts Institute of Technology,
Cambridge, Massachusetts 02139, USA}
\author{Leonardo Banchi}
\affiliation{Department of Physics and Astronomy, University College London, Gower Street,
London WC1E 6BT, UK}
\affiliation{QOLS, Blackett Laboratory, Imperial College London, London SW7 2AZ, UK}

\begin{abstract}
We review recent results on the simulation of quantum channels,
the reduction of adaptive protocols (teleportation stretching),
and the derivation of converse bounds for quantum and private
communication, as established in PLOB [Pirandola, Laurenza,
Ottaviani, Banchi, arXiv:1510.08863]. We start\ by introducing a
general weak converse bound for private communication based on the
relative entropy of entanglement. We discuss how combining this
bound with channel simulation and teleportation stretching, PLOB
established the two-way quantum and private capacities of several
fundamental channels, including the bosonic lossy channel. We then
provide a rigorous proof of the strong converse property of these
bounds by adopting a correct use of the Braunstein-Kimble
teleportation protocol for the simulation of bosonic Gaussian
channels. This analysis provides a full justification of claims
presented in the follow-up paper WTB [Wilde, Tomamichel, Berta,
arXiv:1602.08898] whose upper bounds for Gaussian channels would
be otherwise infinitely large. Besides clarifying contributions in
the area of channel simulation and protocol reduction, we also
present some generalizations of the tools to other entanglement
measures and novel results on the maximum excess noise which is
tolerable in quantum key distribution.
\end{abstract}
\maketitle

\section{Introduction}

In quantum
information~\cite{NielsenChuang,Hayashi,SamRMPm,RMP,HolevoBOOK,sera,AdessoR,hybrid,hybrid2},
the area of quantum and private communications is the subject of
an intense theoretical study, also driven by an increasing number
of experimental implementations. This hectic field ranges from
point-to-point
protocols~\cite{BB84,Ekert,Gisin,DiamantiREV,LM05,Gross,Chris,Twoway,Madsen,GG02,filip-th1,weed1,usenkoTH1,weed2,weed2way,ulrik-entropy,2way2modes2,2way2modes,usenkoREVIEW,1DUsenko,1D-tobias,1way2modes,JosephEXP,jeff1,jeff2,jeff3,jeff4,jeff5,jeff6,HuangCV,JeffLAST,JeffLAST2}
to the development of quantum
repeaters~\cite{Briegel,Rep2,Rep3,Rep4,Rep5,Rep6,Rep7,Rep8,Rep9,Rep10,Rep12,Rep14,Rep15,Rep16,Rep17,Rep18,Peter1,Peter2}%
, untrusted relays~\cite{MDI1,CVMDIQKD,CVMDIQKD2,CVMDIQKD3} and, more
generally, a quantum network or quantum
Internet~\cite{Kimble2008,HybridINTERNET,ref1,ref2,ref4}. In this wide
scenario, it is important to know the fundamental limits imposed by quantum
mechanics, which also serve as benchmarks to test the performance of practical
proposals and new technologies. However, the exploration of the ultimate
limits is not easy, especially when it comes to considering quantum and
private communication protocols which involve feedback strategies, where the
parties may interactively update their quantum systems in a real time fashion.

The most important point-to-point quantum communication scenario involves two
remote parties, Alice and Bob, which are connected by a (memoryless) quantum
channel without pre-sharing any entanglement. By using this channel, the two
parties may achieve various quantum tasks, including the reliable transmission
of qubits, the distillation of entanglement bits (ebits) and, finally, the
communication or generation of secret bits. The most general protocols are
based on transmissions through the quantum channel which are interleaved by
local operations (LO) assisted by unlimited and two-way classical
communication (CC), briefly called adaptive LOCCs.

The first approach to simply quantum communication protocols dates back to
Bennett, DiVincenzo, Smolin and Wootters (BDSW)~\cite{B2main}. These authors
introduced the simulation of discrete-variable (DV) Pauli channels via quantum
teleportation, and exploited this tool to reduce a quantum communication
protocol through a Pauli channel into an entanglement distillation protocol
over mixed isotropic states. This transformation was explicitly discussed for
non-adaptive protocols based on one-way CCs but the extension to two-way CCs
is easy. Since then we have witnessed a number of
advances~\cite{HoroTEL,SougatoBowen,Cirac,Niset,MHthesis,Wolfnotes,Leung}.

Most recently, Pirandola, Laurenza, Ottaviani and Banchi (PLOB)~\cite{PLOB}
generalized these precursory ideas in several ways. First of all, PLOB
introduced the most general form of channel simulation in a communication
scenario, where the quantum channel is replaced by an LOCC (not necessarily
teleportation~\cite{tele,teleCV,telereview}) applied to the input and some
resource state. Furthermore these elements (LOCC and resource state) may be
asymptotic, i.e., defined as limit of suitable sequences. In this way, any
quantum channel can be simulated at any dimension, i.e., both DV channels and
continuous-variable (CV) channels. For instance, this approach allows one to
deterministically simulate the amplitude damping channel, which was impossible
with any of the previous approaches, including the one formulated in
Ref.~\cite{HoroTEL}, whose limitation was due to the use of finite-dimensional
and non-asymptotic LOCCs (see Eq.~(11) in Ref.~\cite{HoroTEL}).

The second advance brought by PLOB was teleportation stretching. This
technique is based on the channel simulation and allows one to re-order an
arbitrary adaptive protocol into a much simpler block version, where the
output state is simply expressed in terms of tensor-product states up to a
single LOCC. This technique works for any channel, at any dimension and for
any type of adaptive protocol, i.e., for any task. In contrast with the BDSW
reduction into entanglement distillation, teleportation stretching maintains
the original task of the protocol, so that adaptive key generation is reduced
into block key generation. This feature is crucial in order to apply the
technique to many different contexts, including the reduction of adaptive
quantum metrology and quantum channel discrimination~\cite{Metro,MetroREVIEW},
and the extension to multi-party protocols~\cite{multipoint} and quantum
networks~\cite{ref1}.

By using teleportation stretching and extending the notion of relative entropy
of entanglement (REE)~\cite{RMPrelent,VedFORMm,Pleniom} from states to
channels, PLOB derived simple single-letter upper bounds for the two-way
quantum and private capacities of an arbitrary quantum channel. In particular,
these capacities were established for dephasing channels, erasure channels
(see also Refs.~\cite{ErasureChannelm,GEWa}), quantum-limited amplifiers, and
bosonic lossy channels. The two-way capacity of the lossy channel, also known
as PLOB bound, closes a long-standing investigation started in
2009~\cite{RevCohINFO,ReverseCAP,TGW}, and finally sets the ultimate limit for
optical quantum communications in the absence of repeaters.

In this manuscript, not only we review these techniques and results, but we
also show some generalizations. We study the general conditions that an
entanglement measure needs to satisfy in order to be used for the derivation
of single-letter upper bounds for two-way assisted capacities. We then
consider a problem which is complementary to that presented in PLOB. Instead
of analyzing the optimal achievable rates in quantum key distribution (QKD),
we investigate the maximum excess noise which is tolerable by QKD\ protocols.
As we will see, this characterization is and remains an open problem.

Finally, we also investigate strong converse properties. In fact, directly
building on the methods described above (channel's REE and teleportation
stretching), a follow-up work by Wilde, Tomamichel and Berta
(WTB)~\cite{WildeFollowup} studied the strong converse property of the upper
bounds established in PLOB. Here we re-consider this later refinement while
fixing its technical mathematical issues. In fact, we show that the strong
converse bounds for bosonic Gaussian channels presented in WTB technically
explode to infinity, due to an imprecise use of the Braunstein-Kimble
(BK)\ protocol for CV teleportation~\cite{teleCV,teleCV2}.

The optimal teleportation simulation of bosonic channels is asymptotic and,
for this reason, must be handled with a careful control on the simulation
error (between the actual and the simulated channel). Such error needs to be
rigorously propagated through the protocol and then bounded via a correct use
of the BK teleportation protocol. While this technique is fully taken into
account in the weak converse bounds of PLOB, it is absent in the derivations
of WTB for bosonic Gaussian channels, whose strong converse bounds become
therefore \textquotedblleft unbounded\textquotedblright.

\subsection*{Structure of the paper}

The paper is organized as follows. In Sec.~II, we define adaptive protocols
and two-way capacities. In Sec.~III, we provide a general weak converse upper
bound for these capacities. To simplify this bound, we describe channel
simulation in Sec.~IV and teleportation stretching in Sec.~V. We combine all
these elements in Sec.~VI to derive single-letter upper bounds and the
formulas for the two-way capacities. In Sec.~VII we specify some of the
results to establish the maximum excess noise which is tolerable in QKD. In
Sec.~VIII, we discuss how the recipe introduced by PLOB is general and can be
formulated for other entanglement measures. Sec.~IX contains a detailed
discussion on the main advances in the field of channel simulation and
protocol reduction before the full generalization in PLOB. Then, in Sec.~X, we
review aspects of WTB\ and we provide a complete proof of its strong converse
claims for bosonic Gaussian channels. Finally, Sec.~XI is for conclusions.

\section{Adaptive quantum protocols and two-way capacities}

\subsection{General formulation and definitions}

\begin{figure}[pth]
\vspace{-2.0cm}
\par
\begin{center}
\vspace{-0.0cm} \includegraphics[width=0.50\textwidth]{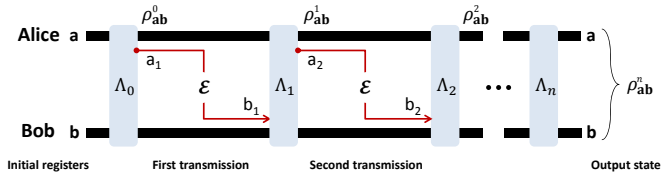}
\vspace{-2.7cm}
\end{center}
\caption{Adaptive protocol through a quantum channel $\mathcal{E}$. Each
transmission $a_{i}\rightarrow b_{i}$ is interleaved by adaptive LOCCs,
$\Lambda_{i-1}$ and $\Lambda_{i}$, applied to the local registers $\mathbf{a}$
and $\mathbf{b}$. After $n$ transmissions, we have a sequence of adaptive
LOCCs $\mathcal{P}=\{\Lambda_{0},\ldots,\Lambda_{n}\}$ characterizing the
protocol and a corresponding output state $\rho_{\mathbf{ab}}^{n}$ for Alice
and Bob.}%
\label{longPIC}%
\end{figure}

Let us start with the description of the most general adaptive protocol for
quantum or private communication over an arbitrary quantum channel
$\mathcal{E}$. We adopt the notation introduced in PLOB, where Alice and Bob
have local registers, $\mathbf{a}$ and $\mathbf{b}$, each composed of a
(countable) number of systems. The adaptive protocol goes as
follows~\cite{PLOB} (see also Fig.~\ref{longPIC} for a schematic).

\begin{itemize}
\item Alice and Bob prepare an initial state $\rho_{\mathbf{ab}}^{0}$ applying
an adaptive LOCC $\Lambda_{0}$ to their registers $\mathbf{a}$ and
$\mathbf{b}$.

\item Alice picks a system $a_{1}\in\mathbf{a}$ and sends it through the
channel $\mathcal{E}$; Bob receives the output system $b_{1}$ which becomes
part of his register, $b_{1}\mathbf{b}\rightarrow\mathbf{b}$. Another adaptive
LOCC $\Lambda_{1}$ is then applied to the local registers, with output
$\rho_{\mathbf{ab}}^{1}$.

\item Then, there is the second transmission $\mathbf{a}\ni a_{2}\rightarrow
b_{2}$ through $\mathcal{E}$, which is followed by another adaptive LOCC
$\Lambda_{2}$. This procedure goes so on for $n$ uses of the channel. The
sequence of adaptive LOCCs $\mathcal{P}=\{\Lambda_{0},\ldots,\Lambda_{n}\}$
characterizes the protocol and provides the output state $\rho_{\mathbf{ab}%
}^{n}$.
\end{itemize}

The output state $\rho_{\mathbf{ab}}^{n}$ is epsilon-close to some target
state $\phi^{n}$ with $nR_{n}$ bits, where $R_{n}$ is the rate. In other
words, we have $||\rho_{\mathbf{ab}}^{n}-\phi^{n}||\leq\varepsilon$ in trace
norm~\cite{Eps}. Depending on the task of the protocol, the target bits are of
different types, e.g., qubits, ebits or private bits. The \textit{generic}
two-way capacity is taking the limit of large $n$ and optimizing over the
adaptive protocols%
\begin{equation}
\mathcal{C}(\mathcal{E}):=\sup_{\mathcal{P}}\lim_{n}R_{n}. \label{def1}%
\end{equation}
If the target $\phi^{n}$ is a maximally-entangled state, then $\mathcal{C}$ is
the two-way entanglement-distribution capacity $D_{2}$. Under two-way CCs,
$D_{2}$ is equal to the quantum capacity $Q_{2}$ (transmission of qubits). If
the target $\phi^{n}$ is a private state~\cite{KD}, then $\mathcal{C}$ is the
secret key capacity $K$ (generation of secret bits), which is equal to the
two-way private capacity $P_{2}$ (private transmission of classical bits).
Since a maximally-entangled state is a specific type of private state,
entanglement distillation is a particular protocol of key distillation and we
may write the hierarchy
\begin{equation}
Q_{2}=D_{2}\leq K=P_{2}. \label{hierarchy}%
\end{equation}

\subsection{Private states}

Let us explain in detail the structure of a private state~\cite{KD}. Decompose
the local registers as $\mathbf{a}=AA^{\prime}$ and $\mathbf{b}=BB^{\prime}$,
where $A$ and $B$ are the local \textquotedblleft key
systems\textquotedblright\ with individual dimension $d_{\mathrm{K}}$, while
$A^{\prime}$ and $B^{\prime}$ are used to protect the key and form the
so-called \textquotedblleft shield system\textquotedblright, whose dimension
$d_{\mathrm{S}}$ is in principle arbitrary (even infinite for CV systems). The
total dimension of the registers is therefore $d=d_{\mathrm{K}}^{2}%
d_{\mathrm{S}}$.\ A generic private state has the form
\begin{equation}
\phi_{AA^{\prime}BB^{\prime}}=U(\Phi_{AB}\otimes\chi_{A^{\prime}B^{\prime}%
})U^{\dagger}, \label{privateSTATEform}%
\end{equation}
where $U$\ is a \textquotedblleft twisting unitary\textquotedblright%
~\cite{KD}, $\chi_{A^{\prime}B^{\prime}}$ is the state of the shield, and
$\Phi_{AB}$ is the maximally entangled state%
\begin{equation}
\Phi_{AB}=\left\vert \Phi\right\rangle _{AB}\left\langle \Phi\right\vert
,~~\left\vert \Phi\right\rangle _{AB}:=d_{\mathrm{K}}^{-1/2}\sum
_{i=0}^{d_{\mathrm{K}}-1}\left\vert i\right\rangle _{A}\left\vert
i\right\rangle _{B}. \label{EPRfinite}%
\end{equation}

By making local measurements on $AB$ and tracing out the shield $A^{\prime
}B^{\prime}$, Alice and Bob share an ideal private state which is completely
factorized from the eavesdropper (Eve), i.e., of the form~\cite{KD}
\begin{equation}
\tau_{AB\mathbf{e}}=\left(  d_{\mathrm{K}}^{-1}\sum_{i=0}^{d_{\mathrm{K}}%
-1}\left\vert i\right\rangle _{A}\left\langle i\right\vert \otimes\left\vert
i\right\rangle _{B}\left\langle i\right\vert \right)  \otimes\tau_{\mathbf{e}%
}, \label{targetidealQKD}%
\end{equation}
with $\tau_{\mathbf{e}}$ is an arbitrary state for Eve's system $\mathbf{e}$.
It is important to note that system $\mathbf{e}$ can also be embedded in an
infinite-dimensional Hilbert space. In fact, even if Alice and Bob employ DV
systems, Eve may resort to hybrid DV-CV interactions with a CV environment
under her control. However, let us also notice that, even if Eve resorts to a
CV environment, its effective Hilbert space will still be finite-dimensional,
just because the minimum purification of Alice and Bob's state needs a DV system.

The shared randomness in the classical systems $A$ and $B$ provides $\log
_{2}d_{\mathrm{K}}$ secret bits. Thus, the $n$-use target state $\phi^{n}$ in
the previous adaptive protocol is such that
\begin{equation}
\log_{2}d_{\mathrm{K}}:=nR_{n}. \label{privv}%
\end{equation}
The local dimension $d_{\mathrm{K}}$ defines the number of secret bits and is
exponential in $n$ for both DV and CV systems. On the other hand, the
dimension $d_{\mathrm{S}}$ of the shield system is not specified and may be
super-exponential in DVs or even infinite in CVs. For DV systems, it is well
known that we may restrict the effective size of $d_{\mathrm{S}}$ to be at
most exponential in $n$. In fact, there is the following result.

\begin{lemma}
[Shield~\cite{Matthias1a,Matthias2a}]\label{LemmaShield}The effective increase
of the shield size $d_{\mathrm{S}}$ is at most exponential in the number $n$
of copies or channel uses, i.e., $\log_{2}d_{\mathrm{S}}\leq\kappa n$ for some
constant $\kappa$. More precisely, for any protocol, we can design an
approximate protocol with the same asymptotic rate while having an at most
exponential increase of $d_{\mathrm{S}}$.
\end{lemma}

The proof is based on the fact that, for any protocol, one can consider an
approximate protocol with the same asymptotic rate but split into $m=n/n_{0}%
$\ identical and independent blocks of size $n_{0}$. These blocks provide $m$
copies which are subject to key distillation via one-way CCs. This
distillation procedure has a classical communication cost (number of bits
exchanged in the CCs) which is linear in the number $m$ of
copies~\cite{DeveWinter}. Using arguments from Ref.~\cite{KD}, this implies
that the shield size $d_{\mathrm{S}}$ increases at most exponentially in
$m<n$. See~\cite{details} for more mathematical details.

Thus, for DV systems, the previous lemma allows us to restrict Eq.~(\ref{def1}%
) to adaptive protocols $\mathcal{P}$ for which the shield size grows at most
exponentially. For CV systems, this lemma can still be used after a suitable
truncation of the Hilbert space, as explained in the next section.

\section{General weak-converse bound\label{review}}

\subsection{Relative entropy of entanglement}

In order to bound the various two-way capacities in Eq.~(\ref{hierarchy}), one
can resort to the REE. Let us recall that the REE of a quantum state $\rho$ is
defined as~\cite{RMPrelent,VedFORMm,Pleniom}%
\begin{equation}
E_{R}(\rho)=\inf_{\sigma_{s}}S(\rho||\sigma_{s}),
\end{equation}
where $\sigma_{s}$ is a separable state and
\begin{equation}
S(\rho||\sigma_{s}):=\mathrm{Tr}\left[  \rho(\log_{2}\rho-\log_{2}\sigma
_{s})\right]
\end{equation}
is the quantum relative entropy~\cite{RMPrelent}. Note that we may also
consider the regularized version%
\begin{equation}
E_{R}^{\infty}(\rho):=\lim_{n}~n^{-1}E_{R}(\rho^{\otimes n})\leq E_{R}(\rho)~.
\end{equation}

Consider now a DV quantum channel $\mathcal{E}$, with Choi matrix
$\rho_{\mathcal{E}}:=\mathcal{I}\otimes\mathcal{E}(\Phi)$, where $\mathcal{I}%
$\ is the identity channel and $\Phi$ is a maximally-entangled state (e.g., a
Bell state for qubits). Then, we may consider the entanglement flux of the
channel as the REE of its Choi matrix~\cite{PLOB}
\begin{equation}
\Phi(\mathcal{E}):=E_{R}(\rho_{\mathcal{E}}). \label{entFFF}%
\end{equation}
This is a measure of the entanglement (REE) which may be transmitted via a
single use of the channel.

\subsection{Extension to asymptotic states}

Let us now extend the definition of REE to asymptotic states. This is a step
which is introduced to simplify the notation in following formulas. Recall
that an asymptotic state $\sigma$ is the limit of a sequence of bona-fide
states $\sigma^{\mu}$, i.e., $\sigma:=\lim_{\mu}\sigma^{\mu}$. This
formulation is very natural for CV\ systems where the maximally-entangled
state $\Phi$ is itself asymptotic. In fact, this is the ideal
Einstein-Podolsky-Rosen (EPR) state which is realized as the limit of two-mode
squeezed vacuum (TMSV) states $\Phi^{\mu}$, i.e., $\Phi:=\lim_{\mu}\Phi^{\mu}%
$. Here the parameter $\mu:=\bar{n}+1/2$ quantifies both the amount of
squeezing (entanglement) between the two modes and the local energy, i.e., the
mean number of thermal photons in each mode. Also note that the Choi matrix of
a bosonic channel $\mathcal{E}$ is an asymptotic state which is defined by the
limit%
\begin{equation}
\rho_{\mathcal{E}}:=\lim_{\mu}\rho_{\mathcal{E}}^{\mu},~~\rho_{\mathcal{E}%
}^{\mu}:=\mathcal{I}\otimes\mathcal{E}(\Phi^{\mu}). \label{ChoiAPPROX}%
\end{equation}

Now, recall that, given two sequences of states $\sigma_{1}^{\mu}$ and
$\sigma_{2}^{\mu}$, such that $||\sigma_{k}^{\mu}-\sigma_{k}||\overset{\mu
}{\rightarrow}0$ for $k=0$ or $1$, we may write the relative entropy between
the limit states $\sigma_{1}$ and $\sigma_{2}$ as the following%
\begin{equation}
S(\sigma_{1}||\sigma_{2})\leq\underset{\mu\rightarrow\infty}{\lim\inf}%
S(\sigma_{1}^{\mu}||\sigma_{2}^{\mu}).
\end{equation}
This is known as the lower semi-continuity of the relative entropy, a property
which is valid at any dimension~\cite[Theorem~11.6]{HolevoBOOK}. Following
this property, we extend the definition of REE\ to an asymptotic state
$\sigma:=\lim_{\mu}\sigma^{\mu}$ as follows~\cite{PLOB}
\begin{equation}
E_{R}(\sigma):=\inf_{\sigma_{s}^{\mu}}\underset{\mu\rightarrow\infty}{\lim
\inf}S(\sigma^{\mu}||\sigma_{s}^{\mu}), \label{REE_weaker}%
\end{equation}
where $\sigma_{s}^{\mu}$ is a sequence of separable states converging in trace
norm, i.e., $||\sigma_{s}^{\mu}-\sigma_{s}||\rightarrow0$ for separable
$\sigma_{s}$. Thanks to Eq.~(\ref{REE_weaker}), we may extend the definition
of entanglement flux in Eq.$~$(\ref{entFFF}) to bosonic channels, so
that~\cite{PLOB}%
\begin{equation}
\Phi(\mathcal{E}):=\inf_{\sigma_{s}^{\mu}}\underset{\mu\rightarrow\infty}%
{\lim\inf}S(\rho_{\mathcal{E}}^{\mu}||\sigma_{s}^{\mu}). \label{EfluxCV}%
\end{equation}

\subsection{Weak-converse upper bound for private communication}

Once we have clarified how REE is generally defined for states and asymptotic
states, including Choi matrices, we may provide the following result, which
bounds the two-way capacities of an arbitrary quantum channel.

\begin{theorem}
[\cite{PLOB}]For any quantum channel $\mathcal{E}$ (at any dimension, finite
or infinite), the generic two-way capacity $\mathcal{C}(\mathcal{E})$ of
Eq.~(\ref{def1}) satisfies the weak converse bound%
\begin{equation}
\mathcal{C}(\mathcal{E})\leq E_{R}^{\bigstar}(\mathcal{E}):=\sup_{\mathcal{P}%
}\underset{n}{\lim}~n^{-1}E_{R}(\rho_{\mathbf{ab}}^{n})~, \label{mainweak}%
\end{equation}
where $\rho_{\mathbf{ab}}^{n}$ is the output of an $n$-use protocol
$\mathcal{P}$.
\end{theorem}

\noindent The first and complete proof of this Theorem first appeared in the
second arXiv version of PLOB back in 2015~\cite{PLOBv2}. It is repeated here
for the sake of completeness so to avoid misinterpretations. Let us start with
the case of DV systems and then we show the extension to CV systems via
truncation arguments.

Assume that the total dimension of Alice's and Bob's registers $\mathbf{a}$
and $\mathbf{b}$ is equal to $d$. Even though these registers may be generally
composed by a countable number of quantum systems, after $n$ uses of the
channel, only a finite number of systems will effectively contribute to the
generation of a secret key. In fact, we know that the target private state
$\phi^{n}$, and the effective output state $\rho_{\mathbf{ab}}^{n}$
($\varepsilon$-close to the target), has dimension $d=d_{\mathrm{K}}%
^{2}d_{\mathrm{S}}$ which is at most exponential in the number of uses $n$
(see Lemma~\ref{LemmaShield} on the \textquotedblleft shield\textquotedblright%
). In other words, any protocol can be replaced with an approximate protocol
with this exponential scaling. Thus, we may write
\begin{equation}
\log_{2}d\leq\alpha nR_{n}, \label{dimm}%
\end{equation}
for some constant $\alpha$. See also Eq.~(21) of Ref.~\cite{PLOBv2}.

Because $\left\Vert \rho_{\mathbf{ab}}^{n}-\phi^{n}\right\Vert \leq
\varepsilon$, we may then write the Fannes-type inequality~\cite{Donaldmain}
\begin{equation}
\left\vert E_{R}(\rho_{\mathbf{ab}}^{n})-E_{R}(\phi^{n})\right\vert
\leq4\varepsilon\log_{2}d+2H_{2}(\varepsilon), \label{main11}%
\end{equation}
where $H_{2}$ is the binary Shannon entropy~\cite{CoverThomas}. Using
Eq.~(\ref{dimm}) and $nR_{n}\leq E_{R}(\phi^{n})$~\cite{KD}, the previous
inequality implies~\cite{PLOBv2}%
\begin{equation}
R_{n}\leq\frac{E_{R}(\rho_{\mathbf{ab}}^{n})+2H_{2}(\varepsilon)}%
{(1-4\varepsilon\alpha)n}.
\end{equation}
Taking the limit for $n\rightarrow\infty$ (asymptotic rate) and $\varepsilon
\rightarrow0$ (weak converse), we derive%
\begin{equation}
\lim_{n}R_{n}\leq\lim_{n}n^{-1}E_{R}(\rho_{\mathbf{ab}}^{n})~.
\end{equation}
Optimizing over all protocols $\mathcal{P}$, we find Eq.~(\ref{mainweak}). It
is clear that, without loss of generality, the optimization in
Eq.~(\ref{mainweak}) can be implicitly reduced to protocols with exponential
scaling of the shield system.

Let us extend the proof to CV systems. Assume that, after the last LOCC
$\Lambda_{n}$, Alice and Bob apply a trace-preserving LOCC $\mathbb{T}_{d}$ so
that the protocol becomes $\mathbb{T}_{d}\circ\mathcal{P}=\{\Lambda
_{0},\Lambda_{1},\cdots,\Lambda_{n},\mathbb{T}_{d}\}$ whose truncated
$d$-dimensional output state $\rho_{\mathbf{ab}}^{n,d}=\mathbb{T}_{d}%
(\rho_{\mathbf{ab}}^{n})$ is $\varepsilon$-close to a DV private state with
$nR_{n,d}$ bits. We may then repeat the previous derivation for DVs, which
here leads to%
\begin{equation}
R_{n,d}\leq\frac{E_{R}(\rho_{\mathbf{ab}}^{n,d})+2H_{2}(\varepsilon
)}{(1-4\varepsilon\alpha)n}. \label{ffff}%
\end{equation}

It is pedantic to say that Lemma~\ref{LemmaShield} still applies. In fact, the
truncated protocol $\mathbb{T}_{d}\circ\mathcal{P}$ can be stopped after
$n_{0}$ uses, and then repeated $m$ times in an i.i.d. fashion, with
$n=n_{0}m$. One-way key distillation is then applied to the $m$ DV output
copies $(\rho_{\mathbf{ab}}^{n_{0},d})^{\otimes m}$. This implies a number of
bits of CC which is linear in $m$ which, in turn, leads to an (at most)
exponential scaling of the shield size $d_{S}$ in $m$. In other words, we may
write $\log_{2}d_{S}(m)\leq\kappa m$ for constant $\kappa$. This automatically
implies $\log_{2}d_{S}(n)\leq\kappa_{n}n$ where $\lim\inf_{n}\kappa_{n}%
=\kappa$, because it is always possible to find sub-sequences $(n_{0}%
,2n_{0},3n_{0},\ldots)$ of $n$ achieving the lower limit $\kappa$. As a
result, we may always impose the condition in Eq.~(\ref{dimm}) for the total
dimension $d$ of the private state, because one can always find sub-sequences
of $n$ that make it valid as a lower limit.

Now, because $\mathbb{T}_{d}$ is a trace-preserving LOCC, we exploit the
monotonicity of the REE
\begin{equation}
E_{R}(\rho_{\mathbf{ab}}^{n,d})\leq E_{R}(\rho_{\mathbf{ab}}^{n}),
\end{equation}
and rewrite Eq.~(\ref{ffff}) as%
\begin{equation}
R_{n,d}\leq\frac{E_{R}(\rho_{\mathbf{ab}}^{n})+2H_{2}(\varepsilon)}%
{(1-4\alpha\varepsilon)n}~.
\end{equation}
For large $n$ and small $\varepsilon$, this\ leads to
\begin{equation}
\lim_{n}R_{n,d}\leq\lim_{n}n^{-1}E_{R}(\rho_{\mathbf{ab}}^{n}).
\end{equation}
An important observation is that the upper bound does no longer depend on $d$.
As a consequence, in the optimization of $R_{n,d}$ over all protocols
$\mathbb{T}_{d}\circ\mathcal{P}$ we can implicitly remove the truncation.
Explicitly, we may write%
\begin{align}
K(\mathcal{E})  &  =\sup_{d}\sup_{\mathbb{T}_{d}\circ\mathcal{P}}\lim
_{n}R_{n,d}\nonumber\label{main2}\\
&  \leq\sup_{\mathcal{P}}\lim_{n}n^{-1}E_{R}(\rho_{\mathbf{ab}}^{n}%
):=E_{R}^{\bigstar}(\mathcal{E}).
\end{align}

\begin{remark}
[Original 2015 proof]The steps of this proof are the same as those
in the original 2015 proof~\cite{PLOBv2}. The truncation argument
is described after Eq.~(23) of Ref.~\cite{PLOBv2}, where we
introduced a cut-off for the total Hilbert space at the output
(which therefore applies to both the key and shield systems).
Under this cut-off, we repeated the derivation for DV systems.
Exactly as here we used the monotonicity of the REE to write an
upper bound which is independent from the truncated dimension.
Exploiting this independence, the cut-off was relaxed in the final
expression following the same reasoning as above. Note that the
published version of PLOB~\cite{PLOB} contains other equivalent
proofs which have been given for the sake of completeness. One of
these proofs is completely independent from the details of the
shield system.
\end{remark}

\subsection{Rebuttal of some unfounded claims}

Unfortunately, our truncation argument has been misunderstood. Believing that
the truncation was not applied to the shield system, an author recently
claimed that the shield size was unbounded in our 2015 proof for CV
channels~\cite{WildeQcrypt}. This is clearly not the case because a truncation
is applied to the total output state (key plus shield system). As a result of
this misunderstanding, this author started to claim \textquotedblleft rigorous
proofs\textquotedblright\ of results in PLOB (e.g., see Ref.~\cite{WildeComm}%
). Not only these claims are unfounded, but also in stark contradiction with
other statements made by the same author~\cite{WildeQIP,ClaimsW}.

We also noticed that Ref.~\cite{WildeQIP} would claim \textquotedblleft full
justification\textquotedblright\ of the \textquotedblleft
statements\textquotedblright\ presented in other works~\cite{ref2,GEWa,TGW}
which are based on the squashed entanglement. Refs.~\cite{ref2,GEWa,TGW} would
be \textquotedblleft wrong\textquotedblright\ because of the potential
unboundedness of the shield size in CV systems. Let us stress that these
proofs are to be considered correct, because it is easily and implicitly
understood that a truncation argument as the one discussed above applies and
reduces the private state to an effective DV state. Such a truncation can then
be released in all the final bounds derived in Refs.~\cite{ref2,GEWa,TGW}.

\section{Simulation of quantum channels}

To simplify the upper bound of Eq.~(\ref{mainweak}) into a single-letter
quantity, PLOB~\cite{PLOB} has devised the technique of teleportation
stretching, which reduces an adaptive protocol (with any communication task)
into a corresponding block protocol (with the same original task). The first
ingredient in this technique is the LOCC simulation of a quantum channel,
which allows one to \textquotedblleft stretch\textquotedblright\ a channel
into a quantum state. The second step is the exploitation of this simulation
in the adaptive protocol, so that all channel transmissions are replaced by a
tensor product of quantum states. Let us start with the review of the first
step, i.e., channel simulation.

\subsection{LOCC simulation of a quantum channel}

For any quantum channel $\mathcal{E}$, we may consider an LOCC simulation.
This consists of an LOCC $\mathcal{T}$ and a resource state $\sigma$ such
that, for any input state $\rho$, the output of the channel can be expressed
as~\cite{PLOB}
\begin{equation}
\mathcal{E}(\rho)=\mathcal{T}(\rho\otimes\sigma). \label{sigma00}%
\end{equation}
A channel $\mathcal{E}$ which is LOCC-simulable with a resource state $\sigma$
as in Eq.~(\ref{sigma00}) is also called \textquotedblleft$\sigma
$-stretchable\textquotedblright~\cite{PLOB}. For the same channel
$\mathcal{E}$ there may be different choices for $\mathcal{T}$ and $\sigma$,
so that the simulation may be optimized depending on the task under study.
Furthermore, the simulation can also be asymptotic. This means that we may
consider sequences of resource states $\sigma^{\mu}$ such that~\cite{SimLOCC}%
\begin{equation}
\mathcal{E}^{\mu}(\rho)=\mathcal{T}(\rho\otimes\sigma^{\mu}). \label{asy1}%
\end{equation}
and define a quantum channel as a point-wise limit%
\begin{equation}
\mathcal{E}(\rho)=\lim_{\mu}\mathcal{E}^{\mu}(\rho). \label{asy2}%
\end{equation}
This can be expressed in terms of the Bures fidelity as%
\begin{equation}
\lim_{\mu}F\left[  \mathcal{E}^{\mu}(\rho),\mathcal{E}(\rho)\right]  =1,
\end{equation}
where $F(\rho,\rho^{\prime}):=\mathrm{Tr}\sqrt{\sqrt{\rho}\rho^{\prime}%
\sqrt{\rho}}$ for states $\rho$ and $\rho^{\prime}$.

A simple criterion that enables us to identify a good LOCC simulation for a
quantum channel is that of teleportation covariance. A quantum channel
$\mathcal{E}$ is said to be teleportation covariant if, for any teleportation
unitary $U$, i.e., Pauli operators in DVs and phase-space displacements in
CVs~\cite{telereview},\ we may write the following%
\begin{equation}
\mathcal{E}(U\rho U^{\dagger})=V\mathcal{E}(\rho)V^{\dagger}~,
\label{stretchability}%
\end{equation}
for some other unitary $V$~\cite{PLOB}. Note that this is a property of many
channels, including Pauli and erasure channels in DVs, and bosonic Gaussian
channels in CVs. Channels with this property are \textquotedblleft
Choi-stretchable\textquotedblright, which means that they can be simulated by
using their Choi matrix as resource state. More precisely, we can state the following

\begin{criterion}
[Tele-covariance/Choi-stretchability]A teleportation-covariant channel
$\mathcal{E}$ is Choi-stretchable via teleportation, i.e., it can be simulated
by teleporting input states $\rho$ over its Choi matrix $\rho_{\mathcal{E}}$.
For a DV channel, this means
\begin{equation}
\mathcal{E}(\rho)=\mathcal{T}(\rho\otimes\rho_{\mathcal{E}}), \label{teleDV}%
\end{equation}
where $\mathcal{T}$ is teleportation. For a CV channel, this means
\begin{equation}
\mathcal{E}(\rho)=\lim_{\mu}\mathcal{E}^{\mu}(\rho),~~\mathcal{E}^{\mu}%
(\rho)=\mathcal{T}(\rho\otimes\rho_{\mathcal{E}}^{\mu}), \label{asymptotics}%
\end{equation}
where $\mathcal{T}$ is the LOCC of a (modified) BK teleportation protocol and
the sequence $\rho_{\mathcal{E}}^{\mu}$ defines the asymptotic Choi matrix for
large $\mu$.
\end{criterion}

\subsection{Error in the simulation of bosonic channels\label{SECbos111}}

In order to better clarify the previous criterion for CV bosonic channels, let
us recall the details of the BK teleportation protocol. In the standard
formulation, the protocol is implemented by using a TMSV state $\Phi^{\mu}$ as
resource state. This means that Alice's input state $\rho_{a}$ and part
$a^{\prime}$ of a shared TMSV state $\Phi_{a^{\prime}B}^{\mu}$ are detected by
a CV\ Bell detection (composed of a balanced beamsplitter whose output ports
are measured by two conjugate homodyne detectors). The complex outcome
$\alpha$ of the Bell detection is communicated to Bob, who applies the
conditional phase-space displacement $D(-\alpha)$ on his mode $B$. In this
way, Bob obtains the output state $\rho_{B}$ which is a teleported version
$\rho_{a}^{\mu}$ of the input one $\rho_{a}$.

Therefore, let us call $\mathcal{T}$\ the BK\ teleportation LOCC, i.e., the
LOs given by the Bell POVM and the conditional displacements, suitably
averaged over all the Bell outcomes. The output state can be written as
\begin{equation}
\rho_{a}^{\mu}=\mathcal{T}(\rho_{a}\otimes\Phi^{\mu})=\mathcal{I}^{\mu}%
(\rho_{a}),
\end{equation}
where $\mathcal{I}^{\mu}$ is the BK teleportation channel. This channel can be
locally (i.e., point-wise) described by an additive-noise Gaussian channel
with added noise~\cite{GerLimited,RicFINITE}
\begin{equation}
\xi=2\mu-\sqrt{4\mu^{2}-1}~.
\end{equation}
As a result, one has the point-wise convergence of the
BK\ protocol~\cite{teleCV,teleCV2}: for any energy-bounded state (i.e., a
`point') $\rho_{a}$, we may write
\begin{equation}
\lim_{\mu}F(\rho_{a}^{\mu},\rho_{a})=1.
\end{equation}

The discussion can be automatically extended to considering ancillary systems,
so that the input can be taken as a bipartite state $\rho_{Aa}$ whose part $a$
is teleported while part $A$ is just subject to the identity channel
$\mathcal{I}_{A}$. In this case, we have the output state
\begin{equation}
\rho_{Aa}^{\mu}=\mathcal{I}_{A}\otimes\mathcal{T}(\rho_{Aa}\otimes\Phi^{\mu
})=\mathcal{I}_{A}\otimes\mathcal{I}^{\mu}(\rho_{Aa}),
\label{teleportedVERSION}%
\end{equation}
and we may write the limit%
\begin{equation}
\lim_{\mu}F(\rho_{Aa}^{\mu},\rho_{Aa})=1.
\end{equation}
We may formulate this limit in an equivalent way. In fact, for any input state
$\rho_{Aa}$\ (or `point'), let us define the corresponding teleportation
infidelity at energy $\mu$ as%
\begin{equation}
\varepsilon_{\text{BK}}(\mu,\rho_{Aa}):=1-F(\rho_{Aa}^{\mu},\rho_{Aa}).
\label{pointBK}%
\end{equation}
Then, we may write the point-wise limit%
\begin{equation}
\lim_{\mu}\varepsilon_{\text{BK}}(\mu,\rho_{Aa})=0. \label{pointBK2}%
\end{equation}

\begin{figure}[pth]
\vspace{-2.0cm}
\par
\begin{center}
\vspace{0cm} \vspace{+1cm} \includegraphics[width=0.50\textwidth]{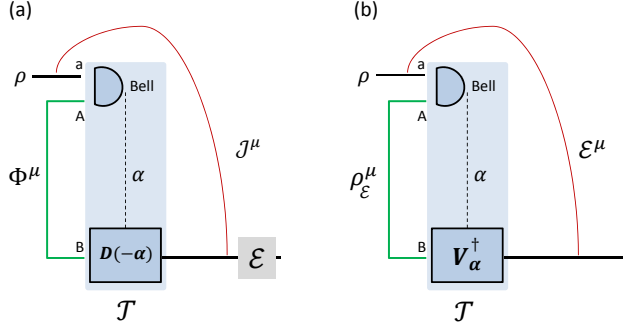}
\vspace{+1cm} \vspace{-2.7cm}
\end{center}
\caption{BK\ protocol and teleportation simulation. (a)~We represent the
standard protocol where a TMSV $\Phi^{\mu}$ and a teleportation LOCC
$\mathcal{T}$ (Bell detection and conditional displacements) are used to
teleport an input state $\rho_{a}$. The input state is equal to $\mathcal{I}%
^{\mu}(\rho_{a})$ where $\mathcal{I}^{\mu}$ is the BK\ teleportation channel.
In general we may consider another bosonic channel $\mathcal{E}$ at the output
so that we have the composition $\mathcal{E}^{\mu}=\mathcal{E}\circ
\mathcal{I}^{\mu}$, as in Eq.~(\ref{asin}). (b)~If the channel is
teleportation-covariant, then we may commute it with the displacement
operators $D(-\alpha)$ up to introducing the modified corrections $V_{\alpha
}^{\dagger}$. As a result the resource state will become the quasi-Choi matrix
$\rho_{\mathcal{E}}^{\mu}$ and the teleportation LOCC $\mathcal{T}$ will be
re-defined over the new correction operators $V_{\alpha}^{\dagger}$. The
channel $\mathcal{E}^{\mu}$ can be represented as $\mathcal{E}^{\mu}(\rho
_{a})=\mathcal{T}(\rho_{a}\otimes\rho_{\mathcal{E}}^{\mu})$ as in
Eq.~(\ref{asin}).}%
\label{BKfig}%
\end{figure}

Consider now a teleportation-covariant bosonic channel $\mathcal{E}$, i.e.,
satisfying~\cite{PLOB}
\begin{equation}
\mathcal{E}\left[  D(\alpha)\rho D(-\alpha)\right]  =V_{\alpha}\mathcal{E}%
(\rho)V_{\alpha}^{\dagger}~,
\end{equation}
for a set of output unitaries $V_{\alpha}$. Then consider its $\mu$-energy
simulation $\mathcal{E}^{\mu}$. This can be realized as in
Eq.~(\ref{asymptotics}) where the resource state is the quasi-Choi matrix
$\rho_{\mathcal{E}}^{\mu}:=\mathcal{I}\otimes\mathcal{E}(\Phi^{\mu})$ and
$\mathcal{T}$ is the LOCC\ of a modified BK\ protocol where the output
correction unitaries are given by $V_{\alpha}^{\dagger}$. For any
energy-constrained input state $\rho_{a}$, the simulated output state can be
written as
\begin{equation}
\mathcal{E}^{\mu}(\rho_{a})=\mathcal{E}\circ\mathcal{I}^{\mu}(\rho
_{a})=\mathcal{T}(\rho_{a}\otimes\rho_{\mathcal{E}}^{\mu}), \label{asin}%
\end{equation}
as also shown in Fig.~\ref{BKfig}. Because $\mathcal{E}^{\mu}=\mathcal{E}%
\circ\mathcal{I}^{\mu}$ and $\mathcal{E}=\mathcal{E}\circ\mathcal{I}$, we may
write
\begin{equation}
\lim_{\mu}F\left[  \mathcal{E}^{\mu}(\rho_{a}),\mathcal{E}(\rho_{a})\right]
\geq\lim_{\mu}F\left[  \mathcal{I}^{\mu}(\rho_{a}),\mathcal{I}(\rho
_{a})\right]  =1,
\end{equation}
i.e., we have the point-wise limit promised in Eq.~(\ref{asymptotics}%
).

This can be extended to the presence of ancillary systems $A$. In fact, given
the teleportation-covariant bosonic channel $\mathcal{I}_{A}\otimes
\mathcal{E}$, its output $\rho_{AB}:=\mathcal{I}_{A}\otimes\mathcal{E}%
(\rho_{Aa})$\ can be simulated by%
\begin{align}
\rho_{AB}^{\mu}  &  :=\mathcal{I}_{A}\otimes\mathcal{E}^{\mu}(\rho_{Aa})\\
&  =\mathcal{I}_{A}\otimes\mathcal{E}\circ\mathcal{I}^{\mu}(\rho
_{Aa})=\mathcal{I}_{A}\otimes\mathcal{T}(\rho_{Aa}\otimes\rho_{\mathcal{E}%
}^{\mu}).
\end{align}
In other words, we may write the point-wise limit%
\begin{equation}
\lim_{\mu}F(\rho_{AB}^{\mu},\rho_{AB})=1. \label{FidLIMIT}%
\end{equation}
Alternatively, we may write this limit in terms of the infidelity%
\begin{align}
\lim_{\mu}\varepsilon_{\text{BK}}(\rho_{AB}^{\mu},\rho_{AB})  &  =0\\
\varepsilon_{\text{BK}}(\rho_{AB}^{\mu},\rho_{AB})  &  :=1-F(\rho_{AB}^{\mu
},\rho_{AB}).
\end{align}

\subsection{Considerations for bosonic Gaussian channels\label{secGAUSS}}

It is very important to note that the previous limit are point-wise and
performed over energy-constrained input states. The situation can be
completely different when, at the input, we also include asymptotic states,
defined as the limit of sequences of states for increasing energy. In fact, in
the presence of an unbounded input alphabet, one needs to take special care on
how the limits are performed. To explore this issue, consider the case of
single-mode bosonic Gaussian channels.

Consider a bosonic mode with vectorial quadrature $\mathbf{\hat{x}}=(\hat
{q},\hat{p})^{T}$ with $[\hat{q},\hat{p}]=i$. Then, consider a single-mode
Gaussian channel $\mathcal{E}$ acting on an input state with mean value
$\mathbf{\bar{x}}=(\bar{q},\bar{p})^{T}$ and covariance matrix (CM)
$\mathbf{V}$. Its action is described by the transformation%
\begin{equation}
\mathbf{\bar{x}\rightarrow\mathbf{T}\bar{x}+d},~~\mathbf{V}\rightarrow
\mathbf{TVT}^{T}+\mathbf{N,} \label{mapGG}%
\end{equation}
where $\mathbf{d}\in R^{2}$ is a displacement, while transmission matrix
$\mathbf{T}$ and the noise matrix $\mathbf{N}$ are $2\times2$ real matrices,
with $\mathbf{N}^{T}=\mathbf{N}\geq0$ and
\begin{equation}
\det\mathbf{N}\geq\left(  \det\mathbf{T}-1\right)  ^{2}. \label{bona_fide_N}%
\end{equation}

Up to input/output unitary transformations, any such channel can be reduced to
a canonical form~\cite{RMP,Caruso,HolevoVittorio,HolevoFORMS} characterized by
zero displacement ($\mathbf{d}=0$) and diagonal matrices $\mathbf{T}$ and
$\mathbf{N}$. Among these forms the phase-insensitive ones are the
thermal-loss channel, the quantum amplifier and the additive-noise Gaussian
channel. These channels can be described by
\begin{equation}
\mathbf{\bar{x}}\rightarrow\sqrt{\eta}\mathbf{\bar{x}},~~\mathbf{V}%
\rightarrow\eta\mathbf{V}+\nu\mathbf{I,} \label{Act}%
\end{equation}
where $\eta\in\mathbb{R}$, $\nu\geq0$, and $\mathbf{I}:=\mathrm{diag}(1,1)$.
In particular, we have the following specifications of Eq.~(\ref{Act}):

\begin{itemize}
\item Thermal-loss channel $\mathcal{E}_{\eta,\bar{n}}$ is characterized by
transmissivity $\eta\in\lbrack0,1]$ and mean thermal number $\bar{n}\geq0$, so
that $\nu=(1-\eta)(\bar{n}+1/2)$. The lossy channel, or pure-loss channel,
corresponds to $\bar{n}=0$.

\item Quantum amplifier $\mathcal{E}_{g,\bar{n}}$ is characterized by gain
$\eta=g>1$ and mean thermal number $\bar{n}\geq0$, so that $\nu=(g-1)(\bar
{n}+1/2)$. A quantum-limited amplifier corresponds to the specific case
$\bar{n}=0$.

\item Additive-noise Gaussian channel $\mathcal{E}_{\xi}$ has transmissivity
$\eta=1$ and additive noise $\nu=\xi\geq0$.
\end{itemize}

\noindent There are other canonical forms which are instead sensitive to
phase. These are the forms in the classes $A_{2}$, $B_{1}$\ and $D$ according
to the terminology introduced by Holevo~\cite{HolevoFORMS}\ and summarized in
Table~I of Ref.~\cite{RMP}. For instance, a $B_{1}$ canonical form is
described by
\begin{equation}
\mathbf{\bar{x}}\rightarrow\mathbf{\bar{x}},~~\mathbf{V}\rightarrow
\mathbf{V}+\mathrm{diag}(0,1),
\end{equation}
so that a vacuum noise unit is added to the momentum quadrature only.

\bigskip

Given a Gaussian channel $\mathcal{E}$ and its simulation $\mathcal{E}^{\mu}$
based on the BK\ protocol, we may consider the corresponding output states
$\rho_{AB}:=\mathcal{I}_{A}\otimes\mathcal{E}(\rho_{Aa})$ and $\rho_{AB}^{\mu
}:=\mathcal{I}_{A}\otimes\mathcal{E}^{\mu}(\rho_{Aa})$ for an
energy-constrained input state $\rho_{Aa}$. We may then write the limit in
Eq.~(\ref{FidLIMIT}). However, the limit in Eq.~(\ref{FidLIMIT}) becomes
ambiguous and problematic if we allow for an energy-unbounded alphabet, which
means to include asymptotic states with diverging energy at the input.

For instance, consider an input sequence of TMSV states $\Phi_{Aa}^{\tilde
{\mu}}$ with increasing squeezing $\tilde{\mu}$. Compute the actual output
$\rho_{AB}(\tilde{\mu})=\mathcal{I}_{A}\otimes\mathcal{E}(\Phi_{Aa}%
^{\tilde{\mu}})$ and the simulated output $\rho_{AB}^{\mu}(\tilde{\mu
}):=\mathcal{I}_{A}\otimes\mathcal{E}^{\mu}(\Phi_{Aa}^{\tilde{\mu}})$ for some
simulation energy $\mu$. It is easy to find Gaussian channels, such as the
identity channel (see Appendix~\ref{App1}) or the $B_{1}$ canonical form (see
Ref.~\cite{PLB}), such that the fidelity tends to zero in the limit of
$\tilde{\mu}\rightarrow\infty$ for any finite $\mu$, i.e.,%
\begin{equation}
\lim_{\tilde{\mu}}F[\rho_{AB}^{\mu}(\tilde{\mu}),\rho_{AB}(\tilde{\mu})]=0~.
\label{opp}%
\end{equation}

By comparing Eqs.~(\ref{FidLIMIT}) and~(\ref{opp}), we see that the joint
limit in $\tilde{\mu}$ (energy of the input) and $\mu$ (energy of the
simulation) is not mathematically defined. This issue can be solved in two ways:

\begin{itemize}
\item Specifying a precise order of the limits, i.e., first in the simulation
energy $\mu$ and then in the input energy $\tilde{\mu}$ or size of the
alphabet. This exploits the fact that the underlying BK teleportation protocol
strongly converges to the identity channel (as we further discuss in
Sec.~\ref{SEC_BK}).

\item Bounding the size of the input alphabet imposing an energy constraint.
For this more elegant solution we need to introduce the energy-constrained
diamond distance. This exploits the fact that the BK teleportation protocol
converges to the identity channel according to the energy-constrained uniform
topology (as we further discuss in Sec.~\ref{diamondDsec}).
\end{itemize}

\subsection{Topologies of convergence in the BK protocol and teleportation
simulation of bosonic channels\label{SEC_BK}}

The previous considerations can be formalized in terms of different topologies
of convergence associated with the BK teleportation protocol. Consider a $\mu
$-energy BK protocol, where a teleportation LOCC $\mathcal{T}$ (Bell plus
conditional displacements) is performed over a TMSV state $\Phi^{\mu}$\ with
finite energy $\mu$. Given an input energy-constrained state $\rho_{Aa}$, we
may write its teleported version $\rho_{Aa}^{\mu}$ as in
Eq.~(\ref{teleportedVERSION}). We also know~\cite{teleCV,teleCV2} that we may
write the point-wise limit of Eq.~(\ref{pointBK2}) for the infidelity
$\varepsilon_{\text{BK}}(\mu,\rho_{Aa}):=1-F(\rho_{Aa}^{\mu},\rho_{Aa})$. By
taking the supremum, it is trivial to write the strong convergence limit%
\begin{equation}
\sup_{\rho_{Aa}}\left[  \lim_{\mu}\varepsilon_{\text{BK}}(\mu,\rho
_{Aa})\right]  =0. \label{conv1}%
\end{equation}
This is also trivially extended to teleportation simulation. In fact, for the
finite-energy simulation $\mathcal{E}^{\mu}$ of a teleportation-covariant
bosonic channel $\mathcal{E}$, we may write the point-wise limit of
Eq.~(\ref{FidLIMIT}) that leads to
\begin{equation}
\sup_{\rho_{Aa}}\left[  \lim_{\mu}\varepsilon_{\text{BK}}(\rho_{AB}^{\mu}%
,\rho_{AB})\right]  =0.
\end{equation}

It is clear, from the reasonings on the order of the limits in
Sec.~\ref{secGAUSS}, that the BK\ protocol does not converge
\textit{uniformly} to the identity channel. In fact, Eq.~(\ref{opp}) written
for $\mathcal{E}=\mathcal{I}$ implies that, for any finite $\mu$, we have%
\begin{equation}
F[\mathcal{I}_{A}\otimes\mathcal{I}^{\mu}(\Phi_{Aa}^{\tilde{\mu}}),\Phi
_{Aa}^{\tilde{\mu}}]\overset{\tilde{\mu}}{\rightarrow}0~, \label{lllkkk}%
\end{equation}
so that $\varepsilon_{\text{BK}}(\mu,\Phi_{Aa}^{\tilde{\mu}})\overset
{\tilde{\mu}}{\rightarrow}1$. Because of this limit, we have%
\begin{equation}
\sup_{\rho_{Aa}}\varepsilon_{\text{BK}}(\mu,\rho_{Aa})=1,~~\text{for any }%
\mu\text{,}%
\end{equation}
and, therefore,%
\begin{equation}
\lim_{\mu}\left[  \sup_{\rho_{Aa}}\varepsilon_{\text{BK}}(\mu,\rho
_{Aa})\right]  =1.
\end{equation}

In diamond distance, this is equivalently to state that%
\begin{equation}
\left\Vert \mathcal{I}-\mathcal{I}^{\mu}\right\Vert _{\diamond}=2,\text{~~for
any }\mu\text{.} \label{IImu}%
\end{equation}
In fact, recall that the diamond distance between two quantum channels
$\mathcal{E}_{1}$ and $\mathcal{E}_{2}$ is defined as%
\begin{equation}
\left\Vert \mathcal{E}_{1}-\mathcal{E}_{2}\right\Vert _{\diamond}:=\sup
_{\rho_{Aa}}\left\Vert \mathcal{I}_{A}\otimes\mathcal{E}_{1}(\rho
_{Aa})-\mathcal{I}_{A}\otimes\mathcal{E}_{2}(\rho_{Aa})\right\Vert ,
\label{diamondDIST}%
\end{equation}
where $\left\Vert \cdot\right\Vert $ is the trace norm. For any two states
$\rho$ and $\rho^{\prime}$, we may then write the Fuchs-van de Graaf
inequality%
\begin{equation}
\left\Vert \rho-\rho^{\prime}\right\Vert \geq2\left[  1-F(\rho,\rho^{\prime
})\right]  . \label{FuchsVan}%
\end{equation}
Therefore, it is easy to see that Eq.~(\ref{lllkkk}) implies Eq.~(\ref{IImu}),
by using Eqs.~(\ref{diamondDIST}) and~(\ref{FuchsVan}).

This non-convergence is also true for the teleportation simulation
of a generic bosonic channel. However, for most of the single-mode
bosonic Gaussian channels, the teleportation simulation uniformly
converges to the channels. In fact, given an arbitrary single-mode
Gaussian channel $\mathcal{E}$ with teleportation simulation
$\mathcal{E}^{\mu}$, we may write~%
\begin{equation}
\lim_{\mu}\left\Vert \mathcal{E}-\mathcal{E}^{\mu}\right\Vert _{\diamond}=0,
\label{uniGAUSS}%
\end{equation}
if and only if its noise matrix $\mathbf{N}$ has full rank, i.e.,
$\mathrm{rank}(\mathbf{N})=2$~\cite{PLB}. For Gaussian channels with
$\mathrm{rank}(\mathbf{N})<2$ and other bosonic channels, we need to replace
the uniform convergence of Eq.~(\ref{uniGAUSS}) with a notion of
bounded-uniform convergence which is based on an energy-constrained version of
the diamond distance.

\subsection{Energy-constrained diamond distance\label{diamondDsec}}

A more useful definition of diamond distance for bosonic channels involves the
introduction of an energy constraint at the input~\cite{PLOB,Shirokov}.
Following PLOB, we impose an energy constraint on the entire input space,
including the ancillas. In fact, consider the following set of
energy-constrained bipartite states
\begin{equation}
\mathcal{D}_{N}:=\{\rho_{Aa}~|~\mathrm{Tr}(\hat{N}\rho_{Aa})\leq N\},
\label{finiteALFA}%
\end{equation}
where $\hat{N}$ is the total number operator associated to the input $a$ and
all the ancillas $A$. One can check that $\mathcal{D}_{N}$ is a compact
set~\cite{HolevoCOMPACT}. Then, for two bosonic channels, $\mathcal{E}_{1}$
and $\mathcal{E}_{2}$, we may define the energy-constrained diamond distance
as%
\begin{equation}
\left\Vert \mathcal{E}_{1}-\mathcal{E}_{2}\right\Vert _{\diamond N}%
:=\sup_{\rho_{Aa}\in\mathcal{D}_{N}}\Vert\mathcal{I}_{A}\otimes\mathcal{E}%
_{1}(\rho_{Aa})-\mathcal{I}_{A}\otimes\mathcal{E}_{2}(\rho_{Aa})\Vert~.
\label{defBBBB}%
\end{equation}

Now, for any bounded alphabet $\mathcal{D}_{N}$ with energy $N$, consider the
energy-constrained\ diamond distance between a (teleportation-covariant)
bosonic channel\ $\mathcal{E}$ and its teleportation simulation $\mathcal{E}%
^{\mu}$. This defines the simulation error%
\begin{equation}
\delta(\mu,N):=\left\Vert \mathcal{E}-\mathcal{E}^{\mu}\right\Vert _{\diamond
N}~. \label{errorAA}%
\end{equation}
Because $\mathcal{D}_{N}$ is compact, the point-wise limit in
Eq.~(\ref{FidLIMIT}) implies the following uniform limit
\begin{equation}
\delta(\mu,N)\overset{\mu}{\rightarrow}0\text{~~\textrm{for any finite }}N,
\label{defBBNN}%
\end{equation}
or, equivalently, we may write%
\begin{equation}
\lim_{\mu}\left[  \sup_{\rho_{Aa}\in\mathcal{D}_{N}}\varepsilon_{\text{BK}%
}(\rho_{AB}^{\mu},\rho_{AB})\right]  =0.
\end{equation}
As a result, when we consider the asymptotic simulation in
Eq.~(\ref{asymptotics}) for an arbitrary bosonic channel, we may consider it
either as a point-wise limit or as a uniform limit while assuming an
energy-constrained alphabet $\mathcal{D}_{N}$ at the input (as in
Ref.~\cite{PLOB}).

\subsection{Finite-resource simulation of Gaussian channels}

Very recently, a different type of simulation has been introduced for Gaussian
channels~\cite{GerLimited,Spyros}. As shown in Ref.~\cite{RicFINITE}, these
simulations are not optimal as the asymptotic ones, but they may still provide
very good approximations of the results in PLOB. According to
Ref.~\cite{GerLimited}, a phase-insensitive Gaussian channel $\mathcal{E}%
_{\eta,\nu}$ can be simulated as follows
\begin{equation}
\mathcal{E}_{\eta,\nu}(\rho)=\mathcal{T}_{\eta}(\rho\otimes\sigma_{\nu}),
\label{Finn}%
\end{equation}
where $\mathcal{T}_{\eta}$ denotes the LOCC of a (modified)\ BK teleportation
protocol with gain $\sqrt{\eta}$~\cite{teleCV}, and $\sigma_{\nu}$ is a
zero-mean two-mode Gaussian state with CM
\begin{equation}
\mathbf{V}(\sigma_{\nu})=\frac{1}{2}\left(
\begin{array}
[c]{cc}%
a\mathbf{I} & c\mathbf{Z}\\
c\mathbf{Z} & b\mathbf{I}%
\end{array}
\right)  , \label{resState}%
\end{equation}
where the elements in the CM\ are equal to~\cite{GerLimited}%
\begin{align}
a  &  =\frac{b+(\eta-1)e^{-2r}}{\eta},~~c=\frac{b-e^{-2r}}{\sqrt{\eta}%
},\nonumber\\
b  &  =\frac{-\left\vert \eta-1\right\vert +\eta e^{2r}+e^{-2r}}%
{-e^{2r}\left\vert \eta-1\right\vert +\eta+1}, \label{Eq3bis}%
\end{align}
and the entanglement parameter $r\geq0$ is connected to the channel parameter
via the relation%
\begin{equation}
\nu=\frac{e^{-2r}}{2}(\eta+1).
\end{equation}

For the specific case of a pure-loss channel, the previous simulation cannot
be used because, for this channel, one has $\nu=(1-\eta)/2$ and, therefore,
$b$ becomes singular in Eq.~(\ref{Eq3bis}). For the pure loss channel, a
finite-resource simulation is just provided by teleporting with gain
$\sqrt{\eta}$ over a Gaussian state with CM~\cite{RicFINITE,Andrea}%
\begin{equation}
\sigma_{\eta}=\left(
\begin{array}
[c]{cc}%
a\mathbf{I} & \sqrt{a^{2}-1/4}\mathbf{Z}\\
\sqrt{a^{2}-1/4}\mathbf{Z} & a\mathbf{I}%
\end{array}
\right)  ,~a=\frac{\eta+1}{2(1-\eta)}.
\end{equation}

\section{Teleportation stretching of adaptive protocols}

\subsection{Stretching with non-asymptotic simulations}

\begin{figure*}[pth]
\vspace{-4.5cm}
\par
\begin{center}
\includegraphics[width=0.98\textwidth]{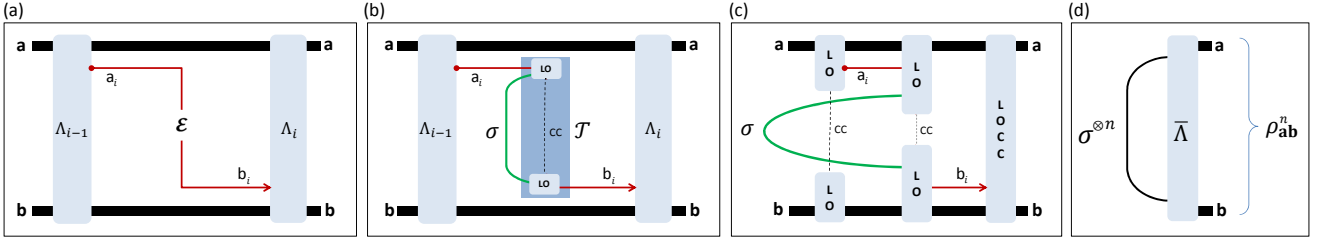} \vspace{-5.0cm}
\end{center}
\caption{Teleportation stretching of an adaptive point-to-point
protocol~\cite{PLOB}. (a)~Consider the generic $i$th transmission through
channel $\mathcal{E}$ between two adaptive LOCCs $\Lambda_{i-1}$ and
$\Lambda_{i}$. (b)~The channel can be simulated by an LOCC $\mathcal{T}$\ and
a resource state $\sigma$. (c) The resource state $\sigma$ is stretched back
in time out of the adaptive LOCCs while $\mathcal{T}$ becomes part of the
LOCCs of the simulated protocol. (d)~By repeating the operation at point~(c)
for all the $n$ transmissions, we accumulate the tensor-product state
$\sigma^{\otimes n}$. All the LOCCs (and also the initial state of the
registers) are collapsed into a single LOCC $\bar{\Lambda}$, which is
trace-preserving after averaging over all measurements. The final result is a
block protocol where the output state $\rho_{\mathbf{ab}}^{n}$ is obtained by
applying $\bar{\Lambda}$ to the resource states $\sigma^{\otimes n}$. This is
the decomposition in Eq.~(\ref{StretchingMAIN}).}%
\label{stretchingPROT}%
\end{figure*}

Thanks to the LOCC-simulation of a quantum channel, we may completely simplify
the structure of an adaptive protocol for quantum/private communication. Let
us start with the simple case of non-asymptotic simulations. Consider an
adaptive protocol with $n$ transmissions over a channel $\mathcal{E}$ which
admits an LOCC-simulation $(\mathcal{T},\sigma)$. Then, we can reduce the
output state $\rho_{\mathbf{ab}}^{n}$ into a tensor-product of resource states
$\sigma^{\otimes n}$ up to a trace-preserving LOCC $\bar{\Lambda}$. In other
words, we may write~\cite{PLOB}
\begin{equation}
\rho_{\mathbf{ab}}^{n}=\bar{\Lambda}\left(  \sigma^{\otimes n}\right)  ~.
\label{StretchingMAIN}%
\end{equation}
As depicted in Fig.~\ref{stretchingPROT}, the procedure goes as follows:

\begin{itemize}
\item Each transmission through $\mathcal{E}$ is replaced by its simulation
$(\mathcal{T},\sigma)$;

\item The resource state $\sigma$ is stretched \textquotedblleft back in
time\textquotedblright\ while $\mathcal{T}$ is included in the LOCCs;

\item All the LOCCs including the register preparation are collapsed into a
single LOCC $\bar{\Lambda}$, which is trace-preserving after averaging over measurements.
\end{itemize}

\subsection{Stretching with asymptotic simulations\label{subsecSTRE}}

Consider now an adaptive protocol with $n$ transmissions over a quantum
channel $\mathcal{E}$ which can be simulated asymptotically by using an LOCC
$\mathcal{T}$ and a sequence of resource states $\sigma^{\mu}$, as in
Eqs.~(\ref{asy1}) and~(\ref{asy2}). The procedure is more involved because we
need to carefully control the propagation of the simulation error from the
channel $\mathcal{E}$ to the final output state $\rho_{\mathbf{ab}}^{n}$.

Let us replace each transmission through $\mathcal{E}$ with an imperfect
channel $\mathcal{E}^{\mu}(\rho):=\mathcal{T}(\rho\otimes\sigma^{\mu})$ based
on a finite-energy resource state $\sigma^{\mu}$. Assuming that, in each $i$th
transmission, the local registers are bounded in energy so that the total
input state $\rho_{\mathbf{a}a_{i}\mathbf{b}}$ belongs to a bounded alphabet
$\mathcal{D}_{N}$, we may write the imperfect simulation with error
$\delta(\mu,N):=\left\Vert \mathcal{E}-\mathcal{E}^{\mu}\right\Vert _{\diamond
N}$ as in Eq.~(\ref{errorAA}). We then need to propagate $\delta(\mu,N)$
throughout the protocol and quantify the trace distance between the actual
output $\rho_{\mathbf{ab}}^{n}:=\rho_{\mathbf{ab}}(\mathcal{E}^{\otimes n})$
and the simulated output $\rho_{\mathbf{ab}}^{n,\mu}:=\rho_{\mathbf{ab}%
}(\mathcal{E}^{\mu\otimes n})$. For any finite $N$, we find~\cite{PLOB}
\begin{equation}
\left\Vert \rho_{\mathbf{ab}}^{n}-\rho_{\mathbf{ab}}^{n,\mu}\right\Vert \leq
n\delta(\mu,N)~. \label{tossll}%
\end{equation}

The proof exploits basic properties of the trace distance. Starting from the
register state $\rho_{\mathbf{ab}}^{0}$, we write
\begin{align}
\rho_{\mathbf{ab}}^{n}  &  =\Lambda_{n}\circ\mathcal{E}\circ\Lambda
_{n-1}\cdots\circ\Lambda_{1}\circ\mathcal{E}(\rho_{\mathbf{ab}}^{0}),\\
\rho_{\mathbf{ab}}^{n,\mu}  &  =\Lambda_{n}\circ\mathcal{E}^{\mu}\circ
\Lambda_{n-1}\cdots\circ\Lambda_{1}\circ\mathcal{E}^{\mu}(\rho_{\mathbf{ab}%
}^{0}),
\end{align}
where we implicitly assume that channels $\mathcal{E}$ and $\mathcal{E}^{\mu}$
are applied to the input system $a_{i}$ in the $i$-th transmission, so that
$\mathcal{E}^{(\mu)}=\mathcal{I}_{\mathbf{a}}\otimes\mathcal{E}_{a_{i}}%
^{(\mu)}\otimes\mathcal{I}_{\mathbf{b}}$. For simplicity, assume $n=2$. We may
apply the \textquotedblleft peeling\textquotedblright\ argument~\cite{PLOB}
\begin{align}
&  \Vert\rho_{\mathbf{ab}}^{2}-\rho_{\mathbf{ab}}^{2,\mu}\Vert\nonumber\\
&  \overset{(1)}{\leq}\Vert\mathcal{E}\circ\Lambda_{1}\circ\mathcal{E}%
(\rho_{\mathbf{ab}}^{0})-\mathcal{E}^{\mu}\circ\Lambda_{1}\circ\mathcal{E}%
^{\mu}(\rho_{\mathbf{ab}}^{0})\Vert\nonumber\\
&  \overset{(2)}{\leq}\Vert\mathcal{E}\circ\Lambda_{1}\circ\mathcal{E}%
(\rho_{\mathbf{ab}}^{0})-\mathcal{E}\circ\Lambda_{1}\circ\mathcal{E}^{\mu
}(\rho_{\mathbf{ab}}^{0})\Vert\nonumber\\
&  +\Vert\mathcal{E}\circ\Lambda_{1}\circ\mathcal{E}^{\mu}(\rho_{\mathbf{ab}%
}^{0})-\mathcal{E}^{\mu}\circ\Lambda_{1}\circ\mathcal{E}^{\mu}(\rho
_{\mathbf{ab}}^{0})\Vert\nonumber\\
&  \overset{(1)}{\leq}\Vert\mathcal{E}(\rho_{\mathbf{ab}}^{0})-\mathcal{E}%
^{\mu}(\rho_{\mathbf{ab}}^{0})\Vert\nonumber\\
&  +\Vert\mathcal{E}[\Lambda_{1}\circ\mathcal{E}^{\mu}(\rho_{\mathbf{ab}}%
^{0})]-\mathcal{E}^{\mu}[\Lambda_{1}\circ\mathcal{E}^{\mu}(\rho_{\mathbf{ab}%
}^{0})]\Vert\nonumber\\
&  \overset{(3)}{\leq}2\Vert\mathcal{E}-\mathcal{E}^{\mu}\Vert_{\Diamond N}~,
\label{casen2}%
\end{align}
where we use the monotonicity of the relative entropy under maps (1), the
triangle inequality (2) and the definition of energy-constrained diamond
distance (3). Generalization to $n\geq2$ provides the result in
Eq.~(\ref{tossll}).

The next step is the stretching of the simulated protocol, i.e., the
decomposition of the state $\rho_{\mathbf{ab}}^{\mu,n}$. By repeating the
steps in Fig.~\ref{stretchingPROT} with $\mathcal{E}^{\mu}$ in the place of
the original channel $\mathcal{E}$, we derive the decomposition $\rho
_{\mathbf{ab}}^{\mu,n}=\bar{\Lambda}\left(  \sigma^{\mu\otimes n}\right)
$\ for a trace-preserving LOCC $\bar{\Lambda}$. For any energy constraint $N$,
we may therefore write%
\begin{equation}
\left\Vert \rho_{\mathbf{ab}}^{n}-\bar{\Lambda}\left(  \sigma^{\mu\otimes
n}\right)  \right\Vert \leq n\delta(\mu,N)~. \label{fromg}%
\end{equation}
For finite energy $N$ and number of uses $n$, we may take the limit of
$\mu\rightarrow\infty$\ and get the asymptotic stretching
\begin{equation}
\left\Vert \rho_{\mathbf{ab}}^{n}-\bar{\Lambda}(\sigma^{\mu\otimes
n})\right\Vert \overset{\mu}{\rightarrow}0. \label{stretchASYm}%
\end{equation}

\section{Single-letter upper bounds}

The most crucial insight of PLOB~\cite{PLOB} has been the combination of the
channel's REE, as expressed by the general weak converse bound in
Eq.~(\ref{mainweak}), with the adaptive-to-block reduction realized by
teleportation stretching, as expressed by Eqs.~(\ref{StretchingMAIN})
and~(\ref{stretchASYm}). This has been the novel recipe that led PLOB to the
computation of extremely simple single-letter upper bounds for all the two-way
capacities of a quantum channel. This entire technique of \textquotedblleft
channel's REE and teleportation stretching\textquotedblright\ has been later
used as a tool in a number of other
works~\cite{Tompaper,multipoint,RicFINITE,ref1} and is at the core of
WTB~\cite{WildeFollowup} and other follow-up papers.

\subsection{Bounds for channels with standard simulations}

Let us start with a quantum channel having a standard (non-asymptotic)
simulation with resource state $\sigma$. Let us compute the REE\ of the output
state $\rho_{\mathbf{ab}}^{n}$ of an adaptive protocol over this channel. By
using the decomposition in Eq.~(\ref{StretchingMAIN}), we derive%
\begin{equation}
E_{R}(\rho_{\mathbf{ab}}^{n})\overset{(1)}{\leq}E_{R}(\sigma^{\otimes
n})\overset{(2)}{\leq}nE_{R}(\sigma)~, \label{toREP}%
\end{equation}
where we use the monotonicity of the REE under trace-preserving LOCCs in~(1),
and its subadditivity over tensor products in~(2). By replacing
Eq.~(\ref{toREP}) in Eq.~(\ref{mainweak}), we then find the single-letter
upper bound~\cite[Theorem~5]{PLOB}
\begin{equation}
\mathcal{C}(\mathcal{E})\leq E_{R}^{\infty}(\sigma)\leq E_{R}(\sigma)~.
\label{UB1}%
\end{equation}
In particular, if the channel $\mathcal{E}$ is teleportation-covariant, it is
Choi-stretchable, and we may write~\cite[Theorem 5]{PLOB}
\begin{equation}
\mathcal{C}(\mathcal{E})\leq E_{R}(\rho_{\mathcal{E}})=\Phi(\mathcal{E}),
\label{UB2}%
\end{equation}
so that the entanglement flux of the channel $\Phi(\mathcal{E})$ bounds all
its two-way assisted capacities. The computation of the single-letter quantity
$\Phi(\mathcal{E})$ is very simple.

\subsection{Formulas for Pauli and erasure channels}

Consider a qubit Pauli channel $\mathcal{P}$ whose action on an input state
$\rho$ is given by%
\begin{equation}
\mathcal{P}(\rho)=p_{0}\rho+p_{1}X\rho X+p_{2}Y\rho Y+p_{3}Z\rho Z,
\end{equation}
where $X$, $Y$, and $Z$ are Pauli operators~\cite{NielsenChuang} and
$\{p_{k}\}$ is a probability distribution. It is easy to check that
$\mathcal{P}$ is teleportation covariant and, therefore, Choi-stretchable.
Computing its entanglement flux $\Phi(\mathcal{P})$, one finds
that~\cite{PLOB}%
\begin{equation}
\mathcal{C}(\mathcal{P})\leq\Phi(\mathcal{P})=1-H_{2}(p_{\max}),
\label{pauliflux}%
\end{equation}
if $p_{\max}:=\max\{p_{k}\}\geq1/2$, while $\Phi=0$ otherwise. The result can
be generalized to arbitrary finite dimension~\cite{PLOB}.

A particular type of Pauli channel is the depolarizing channel $\mathcal{P}%
_{\text{\textrm{depol}}}$, which is defined by
\begin{equation}
\mathcal{P}_{\text{\textrm{depol}}}(\rho)=(1-p)\rho+pI/2, \label{depolQUBITs}%
\end{equation}
for some probability $p$. Specifying Eq.~(\ref{pauliflux}), we
find~\cite{PLOB}%
\begin{equation}
\mathcal{C}(\mathcal{P}_{\text{\textrm{depol}}})\leq\Phi(\mathcal{P}%
_{\text{\textrm{depol}}})=1-H_{2}\left(  3p/4\right)  , \label{depmain}%
\end{equation}
for $p\leq2/3$, while $\Phi=0$ otherwise. Another type of Pauli channel is the
dephasing channel $\mathcal{P}_{\text{\textrm{deph}}}$, defined by
\begin{equation}
\mathcal{P}_{\text{\textrm{deph}}}(\rho)=(1-p)\rho+pZ\rho Z,
\end{equation}
where $p$ is the probability of a phase flip. For this channel, we compute the
entanglement flux~\cite{PLOB}
\begin{equation}
\Phi(\mathcal{P}_{\text{\textrm{deph}}})=1-H_{2}(p)~.
\end{equation}

This upper bound coincides with a lower bound to the capacity which is given
by the one-way distillability of the Choi matrix, i.e., $D_{1}(\rho
_{\mathcal{P}_{\text{\textrm{deph}}}})$. The latter is lower bounded by the
maximum between the coherent~\cite{QC1,QC2} and reverse
coherent~\cite{RevCohINFO,ReverseCAP} information. Because $\Phi
(\mathcal{P}_{\text{\textrm{deph}}})=D_{1}(\rho_{\mathcal{P}%
_{\text{\textrm{deph}}}})$, the dephasing channel is also called
\textquotedblleft distillable\textquotedblright\ and its two-way capacity is
completely determined. We have~\cite{PLOB}
\begin{equation}
\mathcal{C}(\mathcal{P}_{\text{\textrm{deph}}})=1-H_{2}(p)~.
\end{equation}
Note that this also proves $Q_{2}(\mathcal{P}_{\text{\textrm{deph}}%
})=Q(\mathcal{P}_{\text{\textrm{deph}}})$, where the latter was derived in
Ref.~\cite{degradable}.

Consider now an erasure channel which is a non-Pauli channel described by%
\begin{equation}
\mathcal{E}_{\text{\textrm{erase}}}(\rho)=(1-p)\rho+p\left\vert e\right\rangle
\left\langle e\right\vert , \label{erasureDEF}%
\end{equation}
where $p$ is the probability of getting an orthogonal erasure state
$\left\vert e\right\rangle $. This channel is teleportation covariant and also
distillable, therefore we may compute~\cite{PLOB}%
\begin{equation}
\mathcal{C}(\mathcal{E}_{\text{\textrm{erase}}})=\Phi(\mathcal{E}%
_{\text{\textrm{erase}}})=1-p~.
\end{equation}

\begin{remark}
Note that only the $Q_{2}$ of the erasure channel was previously
known~\cite{ErasureChannelm}, so that the novel result here is about the
secret key capacity, i.e., $K(\mathcal{E}_{\text{\textrm{erase}}}%
)=P_{2}(\mathcal{E}_{\text{\textrm{erase}}})=1-p$. Simultaneously with
Ref.~\cite{PLOB}, an independent study of the erasure channel has been
provided in Ref.~\cite{GEWa} which computed the secret key capacity $K$\ from
the squashed entanglement of this channel.
\end{remark}

\subsection{Bounds for channels with asymptotic simulations\label{SEC_CV_asy}}

Consider now a quantum channel $\mathcal{E}$ which is described by an
asymptotic simulation, with an associated sequence of resource states
$\sigma^{\mu}$. For any input alphabet of finite energy $N$ and\ for any
finite number of channel uses $n$, we may write the output of the adaptive
protocol as $\rho_{\mathbf{ab}}^{n}=\lim_{\mu}\bar{\Lambda}(\sigma^{\mu\otimes
n})$ according to the trace norm limit in Eq.~(\ref{stretchASYm}).\ Computing
the REE\ on the (finite-energy) output state $\rho_{\mathbf{ab}}^{n}$ we
find~\cite{PLOB}
\begin{align}
E_{R}(\rho_{\mathbf{ab}}^{n})  &  =\inf_{\sigma_{s}}S(\rho_{\mathbf{ab}}%
^{n}||\sigma_{s})\nonumber\\
&  \overset{(1)}{\leq}\inf_{\sigma_{s}^{\mu}}S\left[  \lim_{\mu}\bar{\Lambda
}(\sigma^{\mu\otimes n})~||~\lim_{\mu}\sigma_{s}^{\mu}\right] \nonumber\\
&  \overset{(2)}{\leq}\inf_{\sigma_{s}^{\mu}}\underset{\mu}{\lim\inf}~S\left[
\bar{\Lambda}(\sigma^{\mu\otimes n})~||~\sigma_{s}^{\mu}\right] \nonumber\\
&  \overset{(3)}{\leq}\inf_{\sigma_{s}^{\mu}}\underset{\mu}{\lim\inf}~S\left[
\bar{\Lambda}(\sigma^{\mu\otimes n})~||~\bar{\Lambda}(\sigma_{s}^{\mu})\right]
\nonumber\\
&  \overset{(4)}{\leq}\inf_{\sigma_{s}^{\mu}}\underset{\mu}{\lim\inf}~S\left(
\sigma^{\mu\otimes n}~||~\sigma_{s}^{\mu}\right) \nonumber\\
&  \overset{(5)}{=}E_{R}(\sigma^{\otimes n}),
\end{align}
where: (1)$~\sigma_{s}^{\mu}$ is a sequence of separable states such that
$\Vert\sigma_{s}-\sigma_{s}^{\mu}\Vert\overset{\mu}{\rightarrow}0$ for
separable $\sigma_{s}$; (2)~we use the lower semi-continuity of the relative
entropy~\cite{HolevoBOOK}; (3)~we use that $\bar{\Lambda}(\sigma_{s}^{\mu})$
are specific types of sequences; (4)~we use the monotonicity of the relative
entropy under $\bar{\Lambda}$; and (5)~we use the definition of REE for
asymptotic states given in Eq.~(\ref{REE_weaker}).

By replacing in Eq.~(\ref{mainweak}), we derive%
\begin{equation}
\mathcal{C}(\mathcal{E}|N)\leq\lim_{n}n^{-1}E_{R}(\sigma^{\otimes n}%
)=E_{R}^{\infty}(\sigma)\leq E_{R}(\sigma)~,
\end{equation}
where we also consider the fact that the capacity is computed assuming an
input alphabet with bounded energy $N$. Because the upper bound does no depend
on $N$, we may extend the result to the supremum and write the final
result~\cite[Theorem 5]{PLOB}%
\begin{equation}
\mathcal{C}(\mathcal{E})=\sup_{N}\mathcal{C}(\mathcal{E}|N)\leq E_{R}^{\infty
}(\sigma)\leq E_{R}(\sigma)~. \label{UBcvm}%
\end{equation}
Exactly as in PLOB, the energy constraint is released at the very end of the calculations.

In particular, for a quantum channel which is teleportation covariant, we may
write the simulation with $\sigma^{\mu}=\rho_{\mathcal{E}}^{\mu}$, i.e.,
considering a Choi sequence. Then, Eq.~(\ref{UBcvm}) becomes
again~\cite[Theorem 5]{PLOB}%
\begin{equation}
\mathcal{C}(\mathcal{E})\leq E_{R}(\rho_{\mathcal{E}})=\Phi(\mathcal{E}),
\end{equation}
where the entanglement flux is defined as in Eq.~(\ref{EfluxCV}). Note that we
may simplify the upper bound by making a specific choice $\tilde{\sigma}%
_{s}^{\mu}$ for the separable sequence $\sigma_{s}^{\mu}$ in
Eq.~(\ref{EfluxCV}), so that
\begin{equation}
\Phi(\mathcal{E})\leq\underset{\mu}{\lim\inf}~S\left(  \rho_{\mathcal{E}}%
^{\mu}~||~\tilde{\sigma}_{s}^{\mu}\right)  . \label{simplerUB}%
\end{equation}

\subsection{Formulas for Gaussian channels}

A single-mode Gaussian channel is teleportation covariant and therefore admits
a teleportation simulation in terms of a Choi sequence $\rho_{\mathcal{E}%
}^{\mu}$. We may bound the generic two-way capacity by using
Eq.~(\ref{simplerUB}) with a suitable separable sequence $\tilde{\sigma}%
_{s}^{\mu}$. Since $\rho_{\mathcal{E}}^{\mu}$ is Gaussian, we may also choose
$\tilde{\sigma}_{s}^{\mu}$ to be Gaussian (see PLOB on how to build this
separable state following ideas in Refs.~\cite{gauss1,gauss2,gauss3}).

The next step is to develop a formula for computing the relative entropy
between two arbitrary Gaussian states. Given $n$ modes with quadratures
$\mathbf{\hat{x}}=(\hat{q}_{1},\ldots,\hat{q}_{n},\hat{p}_{1},\ldots,\hat
{p}_{n})^{T}$, consider the symplectic form
\begin{equation}
\boldsymbol{\Omega}=\left(
\begin{array}
[c]{cc}%
0 & 1\\
-1 & 0
\end{array}
\right)  \otimes I_{n},~
\end{equation}
where $I_{n}$ is the $n\times n$ identity matrix. Using the Gibbs
representation for Gaussian states~\cite{banchiPRL2015}, one can prove the
following~\cite[Theorem~7]{PLOB}

\begin{theorem}
[Relative entropy for Gaussian states]\label{relativeTH}Given two arbitrary
$n$-mode Gaussian states $\rho_{1}(\mathbf{x}_{1},\mathbf{V}_{1})$ and
$\rho_{2}(\mathbf{x}_{2},\mathbf{V}_{2})$, with mean values $\mathbf{x}_{i}$
and CMs $\mathbf{V}_{i}$, their relative entropy is given by
\begin{equation}
S(\rho_{1}||\rho_{2})=-\Sigma(\mathbf{V}_{1},\mathbf{V}_{1})+\Sigma
(\mathbf{V}_{1},\mathbf{V}_{2})~,
\end{equation}
where we have defined
\begin{equation}
\Sigma(\mathbf{V}_{1},\mathbf{V}_{2})=\frac{\ln\det\left(  \mathbf{V}%
_{2}+\frac{i\boldsymbol{\Omega}}{2}\right)  +\mathrm{Tr}(\mathbf{V}%
_{1}\mathbf{G}_{2})+\delta^{T}\mathbf{G}_{2}\delta}{2\ln2}, \label{leoSIG}%
\end{equation}
with $\delta=\mathbf{x}_{1}-\mathbf{x}_{2}$ and $\mathbf{G}_{2}%
=2i\boldsymbol{\Omega}\coth^{-1}(2i\mathbf{V}_{2}\boldsymbol{\Omega})$.
\end{theorem}

\begin{remark}
Note that this formula expresses the relative entropy directly in terms of the
statistical moments of the Gaussian states, without the need of performing the
symplectic diagonalization of the CMs. In fact, Eq.~(\ref{leoSIG}) enables the
use of matrix functions, which are implemented in most numerical and symbolic
software packages. By contrast, a full symplectic diagonalization needs to be
carried out in previous formulations~\cite{Scheelm,Chenmain}. In
Refs.~\cite{Scheelm,Chenmain}, the practical problem is not the computation of
the symplectic spectrum of a CM\ $\mathbf{V}$ (which is relatively easy) but
the derivation of the symplectic matrix $\mathbf{S}$\ performing the
diagonalization $\mathbf{SVS}^{T}=\mathbf{W}$ into the diagonal Williamson
form $\mathbf{W}$~\cite{RMP}. For this matrix $\mathbf{S}$, we know closed
formulas only in very particular cases, e.g., for specific types of two-mode
Gaussian states~\cite{QCB3} as those appearing in problems of quantum
illumination~\cite{Qill1,Qill2,QillGU,Paris} and quantum
reading~\cite{Qread,Qread2}.
\end{remark}

Using Theorem~\ref{relativeTH} for the computation of the relative entropy in
Eq.~(\ref{simplerUB}), PLOB\ established the tightest known upper bounds for
the two-way quantum and private capacities of all single-mode phase
insensitive Gaussian channels. In fact, let us introduce the entropic
function
\begin{equation}
h(x):=(x+1)\log_{2}(x+1)-x\log_{2}x. \label{hEntropyMAIN}%
\end{equation}
Then, we may write the following results~\cite{PLOB}.

\begin{itemize}
\item For a thermal-loss channel $\mathcal{E}_{\eta,\bar{n}}$ with
transmissivity $\eta\in\lbrack0,1]$ and mean thermal number $\bar{n}\geq0$,
one finds%
\begin{equation}
\mathcal{C}(\mathcal{E}_{\eta,\bar{n}})\leq-\log_{2}\left[  (1-\eta)\eta
^{\bar{n}}\right]  -h(\bar{n}), \label{LossUB}%
\end{equation}
for $\bar{n}<\eta/(1-\eta)$, while $\mathcal{C}=0$ otherwise.

\item For a quantum amplifier $\mathcal{E}_{g,\bar{n}}$ with gain $g>1$ and
mean thermal number $\bar{n}\geq0$, one has%
\begin{equation}
\mathcal{C}(\mathcal{E}_{g,\bar{n}})\leq\log_{2}\left(  \dfrac{g^{\bar{n}+1}%
}{g-1}\right)  -h(\bar{n}),
\end{equation}
for $\bar{n}<(g-1)^{-1}$, while $\mathcal{C}=0$ otherwise

\item For an additive-noise Gaussian channel $\mathcal{E}_{\xi}$ with additive
noise $\xi\geq0$, one writes%
\begin{equation}
\mathcal{C}(\mathcal{E}_{\xi})\leq\frac{\xi-1}{\ln2}-\log_{2}\xi,
\end{equation}
for $\xi<1$, while $\mathcal{C}=0$ otherwise.
\end{itemize}

More strongly, for a bosonic lossy channel $\mathcal{E}_{\eta}$\ with
transmissivity $\eta$, PLOB showed that the upper bound $\Phi(\mathcal{E}%
_{\eta})$ coincides with the lower bound $D_{1}(\rho_{\mathcal{E}_{\eta}})$,
the latter being already known from past computations using the reverse
coherent information~\cite{ReverseCAP}. Therefore, a lossy channel is
distillable and the two-way capacities are all equal ($D_{2}=Q_{2}=K=P_{2}$)
and given by~\cite{PLOB}
\begin{equation}
\mathcal{C}(\mathcal{E}_{\eta})=-\log_{2}(1-\eta)~. \label{formCloss}%
\end{equation}

In particular, the secret-key capacity $K$\ of the lossy channel gives the
maximum rate achievable by point-to-point QKD protocols. At high loss
$\eta\simeq0$, one finds the optimal rate-loss scaling $K\simeq1.44\eta$
secret bits per channel use. This result is known as repeaterless or PLOB
bound, and establishes the exact benchmark that a quantum repeater must
surpass in order to be effective.

This result also proves the strict separation $Q_{2}(\mathcal{E}_{\eta
})>Q(\mathcal{E}_{\eta})$, where $Q$ is the unassisted quantum
capacity~\cite{QC1,QC2}. It is then interesting to note that the capacity in
Eq.~(\ref{formCloss}) coincides with the maximum discord~\cite{RMPdiscord}
that can be distributed through the lossy channel, supporting the operational
interpretation of discord as a resource for quantum
cryptography~\cite{DiscordQKD}. One can also check, using the tools in
Ref.~\cite{OptimalDIS}, that this discord corresponds to Gaussian
discord~\cite{GerryDa,ParisDa}.

In conclusion, a quantum-limited amplifier $\mathcal{E}_{g}$ with gain $g>1$
is also distillable, i.e., $\Phi(\mathcal{E}_{g})=D_{1}(\rho_{\mathcal{E}_{g}%
})$. As a result, all the two-way capacities are equal and given
by~\cite{PLOB}
\begin{equation}
C(\mathcal{E}_{g})=-\log_{2}\left(  1-g^{-1}\right)  ~. \label{Campli}%
\end{equation}
In particular, this also proves that $Q_{2}(\mathcal{E}_{g})$ coincides with
the unassisted quantum capacity $Q(\mathcal{E}_{g})$~\cite{HolevoWerner,Wolf}.

\subsection{Amplitude damping channel}

The amplitude damping channel is a very important model of decoherence in spin
chains and networks~\cite{Bose1,Bose2}, especially when we consider the
transfer of quantum information, e.g., in a quantum chip architecture. Despite
this, the inherent asymmetry of this channel makes it the hardest to study. In
the qubit computational basis $\{\left\vert 0\right\rangle ,\left\vert
1\right\rangle \}$, the action of this channel is expressed by
\begin{equation}
\mathcal{E}_{\text{\textrm{damp}}}(\rho)=%
{\textstyle\sum\nolimits_{i=0,1}}
A_{i}\rho A_{i}^{\dagger},
\end{equation}
where $p$ is the damping probability and
\begin{equation}
A_{0}:=\left\vert 0\right\rangle \left\langle 0\right\vert +\sqrt
{1-p}\left\vert 1\right\rangle \left\langle 1\right\vert ,~A_{1}:=\sqrt
{p}\left\vert 0\right\rangle \left\langle 1\right\vert .
\end{equation}

One can check that $\mathcal{E}_{\text{\textrm{damp}}}$ is not
teleportation-covariant. However, it is still LOCC simulable thanks to the
decomposition
\begin{equation}
\mathcal{E}_{\text{\textrm{damp}}}=\mathcal{E}_{\text{\textrm{CV}}%
\rightarrow\text{\textrm{DV}}}\circ\mathcal{E}_{\eta(p)}\circ\mathcal{E}%
_{\text{\textrm{DV}}\rightarrow\text{\textrm{CV}}},
\end{equation}
where:

\begin{itemize}
\item $\mathcal{E}_{\text{\textrm{DV}}\rightarrow\text{\textrm{CV}}}$
teleports the spin qubit into a single-rail bosonic qubit~\cite{telereview};

\item $\mathcal{E}_{\eta(p)}$ is a lossy channel with transmissivity
$\eta(p):=1-p$;

\item $\mathcal{E}_{\text{\textrm{CV}}\rightarrow\text{\textrm{DV}}}$
teleports the single-rail qubit back to the original qubit.
\end{itemize}

For this reason, $\mathcal{E}_{\text{\textrm{damp}}}$ is stretchable into the
asymptotic Choi matrix of the lossy channel $\mathcal{E}_{\eta(p)}$ by means
of a simulating LOCC which combines the local maps $\mathcal{E}%
_{\text{\textrm{CV}}\rightarrow\text{\textrm{DV}}}$ and $\mathcal{E}%
_{\text{\textrm{DV}}\rightarrow\text{\textrm{CV}}}$ with the BK protocol. In
this way, PLOB showed that
\begin{equation}
\mathcal{C}(\mathcal{E}_{\text{\textrm{damp}}})\leq\min\{1,-\log_{2}p\}.
\label{C2dampbb}%
\end{equation}

Let us notice that squashed entanglement can beat this upper bound as shown in
PLOB and Ref.~\cite{GEWa}. The REE bound in Eq.~(\ref{C2dampbb}) is very
simple but performs well only in the regime of high damping ($p>0.9$).
Finally, notice that the amplitude damping proves the need of a
dimension-\textit{independent} theory for channel simulation even if we
restrict ourself to DV channels.

\section{Maximum tolerable noise in quantum key distribution}

In this section we provide a study which complements the one in
Ref.~\cite[Figure~6]{PLOB}, where we plotted the optimal key rate versus
distance of several QKD protocols, in comparison with the PLOB bound. Here we
study the optimal security thresholds which are achieved by setting the key
rates equal to zero.

Consider a thermal-loss channel $\mathcal{E}_{\eta,\bar{n}}$ with
transmissivity $\eta$ and mean thermal number $\bar{n}$. From the variance
parameter $\omega=\bar{n}+1/2$, we define the so-called \textquotedblleft
excess noise\textquotedblright\ $\varepsilon$ of the channel by setting
\begin{equation}
\omega=\frac{1}{2}+\frac{\eta\varepsilon}{1-\eta}, \label{exbb}%
\end{equation}
which leads to%
\begin{equation}
\varepsilon=\eta^{-1}(1-\eta)\bar{n}. \label{excessFORLL}%
\end{equation}
For any protocol, we may write an optimal rate in terms of these channel
parameters, i.e., $R=R(\eta,\varepsilon)$. The security threshold is then
achieved by setting $R=0$ which provides the maximum tolerable excess noise as
a function of the transmissivity, i.e., $\varepsilon=\varepsilon(\eta)$. Now
the crucial question is the following: \textit{What is the maximum excess
noise that is tolerable in QKD? I.e., optimizing over all QKD protocols?}

\begin{figure}[ptb]
\vspace{0.2cm}
\par
\begin{center}
\includegraphics[width=0.40\textwidth]{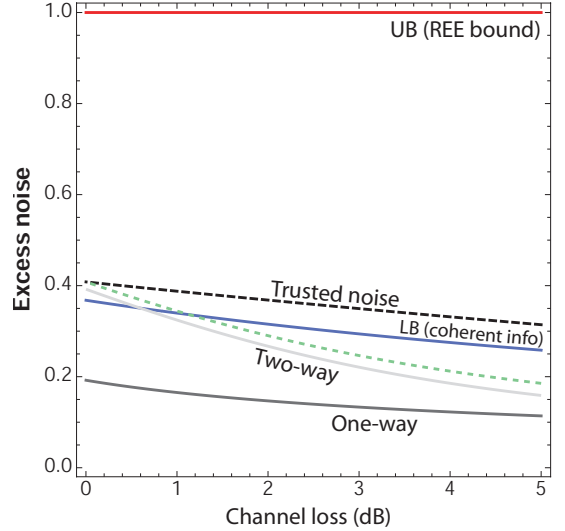}
\end{center}
\par
\vspace{-0.5cm}\caption{Security thresholds in terms of maximum tolerable
excess noise $\varepsilon$ as a function of the loss in the channel (dB).
Protocols are secure below their corresponding thresholds. The red line is the
upper bound $\varepsilon_{\text{UB}}=1$ coming from the entanglement flux (REE
bound) of the thermal-loss channel~\cite{PLOB}. The blue line is the lower
bound $\varepsilon_{\text{LB}}$ computed from the reverse coherent
information~\cite{RevCohINFO,ReverseCAP}. The black dashed line is the
security threshold which is obtained from the key-rate of Eq.~(\ref{rate}),
for the one-way trusted noise protocol described in the text. This is an
improved lower bound, but still far to close the gap with the upper bound.
Finally, we also show the security thresholds corresponding to the one-way
no-switching protocol~\cite{ChrisQKD} and the two-way protocols~\cite{Twoway}
with coherent states (solid line) and largely-thermal states~\cite{PRAthermal}
(green dashed line).}%
\label{thresholds}%
\end{figure}

It is easy to write an upper bound to the security threshold associated with
the secret key capacity of the thermal-loss channel. In fact, from
Eq.~(\ref{LossUB}), we see that $K(\mathcal{E}_{\eta,\bar{n}})=0$, corresponds
to the entanglement-breaking value $\bar{n}=\eta/(1-\eta)$. By replacing in
Eq.~(\ref{excessFORLL}), we find that the maximum tolerable excess noise is
upper-bounded by $\varepsilon_{\text{UB}}=1$ for any value of the
transmissivity $\eta$. For the lower bound, we may consider the maximum key
rate achievable by using the reverse coherent information~\cite{RevCohINFO},
which is equal to~\cite{ReverseCAP}%
\begin{equation}
R_{\mathrm{LB}}=-\log_{2}(1-\eta)-s\left(  \omega\right)  ~, \label{LBrev}%
\end{equation}
where
\begin{equation}
s(x):=\left(  x+\frac{1}{2}\right)  \log_{2}\left(  x+\frac{1}{2}\right)
-\left(  x-\frac{1}{2}\right)  \log_{2}\left(  x-\frac{1}{2}\right)  .
\end{equation}
Using Eq.~(\ref{exbb}) in Eq.~(\ref{LBrev}), we may numerically compute
$R_{\mathrm{LB}}(\eta,\varepsilon)=0$ and find the lower bound $\varepsilon
_{\text{LB}}$. As we can see from Fig.~\ref{thresholds}, there is a huge gap
between $\varepsilon_{\text{LB}}$ and $\varepsilon_{\text{UB}}$.

Can we reduce this gap? From Refs.~\cite{RaulQKD,ReverseCAP}, we know that the
use of trusted noise at the receiver station may improve the performance of a
one-way CV-QKD protocol performed in reverse reconciliation. Such an idea has
been also explored in a recent work~\cite{OTT-SPIE}. Both
Refs.~\cite{ReverseCAP,OTT-SPIE} show that rate associated with the reverse
coherent information can be beaten by a CV-QKD\ protocol based on trusted
noise when non-zero excess noise is present in the channel. Here we show an
equivalent CV-QKD\ protocol which outperforms the security threshold
$\varepsilon_{\text{LB}}$ associated with the reverse coherent information.

The protocol consists of Alice preparing Gaussian-modulated squeezed states,
e.g., by homodyning one part of TMSV states in her hands. It is easy to see
that, for a TMSV state with parameter $\mu$, the local homodyne in $\hat{q}$
on one mode projects the other mode into a displaced $q$-squeezed state with
variance $\mu^{-1}$. Alice randomly switches between $q$- and $p$-squeezed
states following Ref.~\cite{CerfQKD}. The squeezed states are sent through the
thermal-loss channel whose output is measured by Bob. Before detection Bob
applies an additive noise Gaussian channel $\mathcal{E}_{\xi}$, so that the
output quadratures are transformed according to $\mathbf{\hat{x}}%
\rightarrow\mathbf{\hat{x}}+\zeta$, where $\zeta$ is a classical Gaussian
variable with variance $\xi\geq0$. Then, he performs homodyne detection,
switching between the measurement of the $\hat{q}$ and $\hat{p}$ quadrature.

After the parties reconcile their bases, perform error correction and privacy
amplification, they will share an asymptotic key rate $R=(I_{\text{AB}}%
-\chi_{\text{BE}})/2$, where $I_{\text{AB}}$ is Alice and Bob's mutual
information (ignoring the basis reconciliation), and $\chi_{\text{BE}}$
is\ the corresponding Eve's Holevo information on Bob's outcomes. The factor
$1/2$ accounts for the basis reconciliation. After some algebra, we compute%
\begin{align}
I_{\text{AB}}  &  =\frac{1}{2}\log_{2}\frac{\eta\mu+(1-\eta)\omega+\xi}%
{\eta\mu^{-1}+(1-\eta)\omega+\xi}\label{I}\\
&  \overset{\mu}{\rightarrow}\frac{1}{2}\log_{2}\frac{\eta\mu}{(1-\eta
)\omega+\xi}~,
\end{align}
and, for large Gaussian modulation ($\mu\gg1/2$), we get%
\begin{equation}
\chi_{\text{BE}}=\frac{1}{2}\log_{2}\frac{(1-\eta)\eta\mu}{\omega+\xi(1-\eta
)}+s(\omega)-s\left(  \bar{\nu}\right)  , \label{chi}%
\end{equation}
where the symplectic eigenvalue $\bar{\nu}$ is given by%
\begin{equation}
\bar{\nu}=\sqrt{\frac{\omega\left[  1+4\omega\xi(1-\eta)\right]  }%
{4[\omega+\xi(1-\eta)]}}.
\end{equation}

Therefore, from Eqs.~(\ref{I}) and~(\ref{chi}), we compute the asymptotic and
high-modulation rate%
\begin{equation}
R=\frac{1}{4}\log_{2}\frac{\omega+\xi(1-\eta)}{(1-\eta)\left[  (1-\eta
)\omega+\xi\right]  }+\frac{s\left(  \bar{\nu}\right)  -s(\omega)}{2}.
\label{rate}%
\end{equation}
The previous rate is a function of the main parameters, i.e., $R=R(\eta
,\varepsilon,\xi)$. Setting $R=0$, we derive the threshold $\varepsilon
=\varepsilon(\eta,\xi)$ which is maximal in the limit of large trusted noise
$\xi\gg0$. The limit $\varepsilon_{\infty}:=\lim_{\xi}\varepsilon(\eta,\xi)$
beats $\varepsilon_{\text{LB}}$ as shown in Fig.~\ref{thresholds}, therefore
establishing the best-known lower bound. Unfortunately, this is still far from
$\varepsilon_{\text{UB}}$ so that it remains an open problem to establish the
maximum value of tolerable excess noise in QKD.

\begin{remark}
Note that the previous protocol can be implemented in a coherent fashion,
where Alice distributes TMSV states whose her kept modes and Bob's output
modes (from the channel) are stored in quantum memories. The parties may then
agree to perform a joint random sequence of $q$- and $p$- homodyne detections,
so that no basis reconciliation is needed. In this way they can reach a key
rate which is the double of the one in Eq.~(\ref{rate}). This coherent
protocol is equivalent to the one described in Ref.~\cite{OTT-SPIE}, where the
parties use quantum memories and the trusted noise is created by a beam
splitter of transmissivity $\eta_{d}$ mixing the output with a thermal mode
with variance $\gamma$. One can check that the rates are equal by setting%
\begin{equation}
\xi=\frac{1-\eta_{d}}{\eta_{d}}\gamma.
\end{equation}

\end{remark}

\begin{remark}
As one can check, if we set Bob's trusted noise to zero ($\xi=0$), then the
rate in Eq.~(\ref{rate}) becomes equal to half the rate $R_{\mathrm{LB}}$ in
Eq.~(\ref{LBrev}). Assuming a coherent implementation as discussed in the
previous remark, then the rate becomes equal to the rate $R_{\mathrm{LB}}$ in
Eq.~(\ref{LBrev}). Taking the limit for low loss $\eta\simeq0$ and low noise
$\omega\simeq1$, i.e., low thermal photon number $\bar{n}\simeq0$, one may
derive the expansion
\begin{equation}
R_{\mathrm{LB}}\simeq(\eta-\bar{n})\log_{2}e+\bar{n}\log_{2}\bar{n},
\end{equation}
which is the rate studied in Ref.~\cite{Lasota}. Note that Ref.~\cite{Lasota}
also investigated the use of trusted noise at Bob's side in order to increase
the security threshold of the basic squeezed-state protocol~\cite{CerfQKD}.
\end{remark}

For comparison, in Fig.~\ref{thresholds}, we also show the optimal threshold
of the one-way protocol based on Gaussian-modulated coherent states and
heterodyne detection~\cite{ChrisQKD}, whose ideal reverse reconciliation rate
is given by the formula%
\begin{align}
R  &  =\log_{2}\frac{2}{e}\frac{\eta}{\left(  1-\eta\right)  \left[
\eta+2\omega\left(  1-\eta\right)  +1\right]  }\nonumber\\
&  +s\left[  \frac{1+2\omega\left(  1-\eta\right)  }{2\eta}\right]  -s\left(
\omega\right)  .
\end{align}
Then, in Fig.~\ref{thresholds}, we also show the optimal threshold of the
two-way protocol~\cite{Twoway} which is based on the Gaussian modulation of
thermal states (with variance $V_{0}$) and homodyne detection at the output.
Its ideal reverse reconciliation rate is given by~\cite{PRAthermal}%
\begin{equation}
R(V_{0})=\frac{1}{2}\log_{2}\frac{\eta^{2}V_{0}+\omega+\eta^{3}\left(
\omega-V_{0}\right)  }{\left(  1-\eta\right)  \left[  \left(  1-\eta
^{2}\right)  \omega+\eta V_{0}\right]  }+s\left(  \bar{\nu}_{2}\right)
-s\left(  \omega\right)  ,
\end{equation}
where%
\begin{equation}
\bar{\nu}_{2}=\sqrt{\frac{\omega\left[  1+4\eta^{2}V_{0}\omega+\eta
^{3}(1-4\omega V_{0})\right]  }{4\left[  \eta^{2}V_{0}+\omega+\eta^{3}%
(\omega-V_{0})\right]  }}.
\end{equation}
In particular, we consider the limit of coherent states ($V_{0}=1/2$) and that
of largely-thermal states ($V_{0}\gg1$).

\section{General methodology}

Here we discuss how the methodology devised in PLOB for point-to-point
protocols (and extended in Ref.~\cite{ref1} to end-to-end protocols) can be
applied to any entanglement measure $E$ with suitable properties. For
simplicity, here we start considering DV systems. Then we extend the arguments
to CV systems via truncation.

\subsection{Main ingredients}

Assume that $E$ is an entanglement measure that satisfies the following conditions:

\begin{description}
\item[(1) Normalization.] For a target state $\phi^{n}$ encoding $nR_{n}$ bits
(e.g., ebits or private bits), we have
\begin{equation}
E(\phi^{n})\geq nR_{n}. \label{norma}%
\end{equation}

\item[(2) Continuity.] For $d$-dimensional $\rho$ and $\sigma$ such that
$\left\Vert \rho-\sigma\right\Vert \leq\varepsilon$, we have the Fannes-type
inequality
\begin{equation}
\left\vert E(\rho)-E(\sigma)\right\vert \leq g(\varepsilon)\log_{2}%
d+h(\varepsilon), \label{continuity}%
\end{equation}
where $g$, $h$ are regular functions going to zero in $\varepsilon$.

\item[(3) Monotonicity.] For any trace-preserving LOCC $\bar{\Lambda}$, we may
write the data processing inequality%
\begin{equation}
E\left[  \bar{\Lambda}(\rho)\right]  \leq E(\rho)~. \label{monoPROP}%
\end{equation}

\item[(4) Subadditivity.] For any $\rho$ and $\sigma$, we may write%
\begin{equation}
E(\rho\otimes\sigma)\leq E(\rho)+E(\sigma)~, \label{subPROP}%
\end{equation}
so that the regularization satisfies
\begin{equation}
E^{\infty}(\rho):=\lim_{n}n^{-1}E(\rho^{\otimes n})\leq E(\rho)~.
\end{equation}

\end{description}

It is clear that these properties are satisfied by the REE $E_{\text{R}}%
$~\cite{RMPrelent} with the specific choice
\begin{equation}
g(\varepsilon)=4\varepsilon,~h(\varepsilon)=2H_{2}(\varepsilon),
\end{equation}
where $H_{2}$ is the binary Shannon entropy. They are also satisfied by the
squashed entanglement $E_{\text{sq}}$~\cite{squash} with
\begin{equation}
g(\varepsilon)=16\sqrt{\varepsilon},~h(\varepsilon)=2H_{2}(2\sqrt{\varepsilon
}).
\end{equation}

Consider now an arbitrary adaptive protocol $\mathcal{P}$ between two users,
Alice and Bob. This protocol may be point-to-point over a quantum
channel~\cite{PLOB} or an end-to-end protocol along a repeater chain or within
a quantum network~\cite{ref1}. After $n$ uses, assume that Alice and Bob's
output is $\varepsilon$ close to a $d$-dimensional target state $\phi^{n}$. By
applying Eqs.~(\ref{norma}) and~(\ref{continuity}), we derive%
\begin{equation}
R_{n}\leq\frac{E(\rho_{\mathbf{ab}}^{n})+g(\varepsilon)\log_{2}d+h(\varepsilon
)}{n}. \label{main}%
\end{equation}

Now assume the following property for the target state, which is certainly
true for a maximally entangled state and also for a private state.

\begin{description}
\item[(5) Exponential size.] The effective total dimension of the target state
grows at most exponentially
\begin{equation}
d\leq2^{\alpha_{n}n}, \label{sizePROP}%
\end{equation}
where $\lim\inf_{n}\alpha_{n}=\alpha$ for constant $\alpha$. In particular,
for a maximally-entangled state $\alpha_{n}=\alpha=1$.
\end{description}

Combining Eq.~(\ref{main}) and~(\ref{sizePROP}), and taking the limit for
large $n$ we derive%
\begin{align}
\lim_{n}R_{n}  &  \leq\lim_{n}\frac{E(\rho_{\mathbf{ab}}^{n})}{n}+\underset
{n}{\lim\inf}g(\varepsilon)\alpha_{n}+\lim_{n}\frac{h(\varepsilon)}{n}\\
&  =\lim_{n}n^{-1}E(\rho_{\mathbf{ab}}^{n})+g(\varepsilon)\alpha~.
\end{align}
Then, by taking the limit for small $\varepsilon$, we derive the weak converse
bound%
\begin{equation}
\lim_{n}R_{n}\leq\lim_{n}n^{-1}E(\rho_{\mathbf{ab}}^{n})~. \label{UB2b}%
\end{equation}
Finally, consider the optimization over all protocols $\mathcal{P}$ (more
precisely, over the equivalent class that satisfies the property in
Eq.~(\ref{sizePROP}) on the exponential size). This leads to%
\begin{equation}
\mathcal{C}:=\sup_{\mathcal{P}}\lim_{n}R_{n}\leq\sup_{\mathcal{P}}\lim
_{n}n^{-1}E(\rho_{\mathbf{ab}}^{n})~, \label{toSImplify}%
\end{equation}
where $\mathcal{C}$ may be a two-way assisted capacity over a quantum channel,
or an end-to-end capacity of a repeater chain or a quantum network.

The next step is to simplify the upper bound in Eq.~(\ref{toSImplify}) to a
single-letter quantity via a suitable decomposition of the output state. For
simplicity, restrict the analysis to point-to-point protocols over a quantum
channel (generalization to repeaters and networks goes along the lines of
Ref.~\cite{ref1} and requires the introduction of additional tools from
network information theory). We know that we have the following powerful
tool~\cite{PLOB}.

\begin{description}
\item[(6) Teleportation stretching.] Simulating a channel $\mathcal{E}$ with a
resource state $\sigma$, we may re-organize any point-to-point
(generally-adaptive) protocol in a block form so as to decompose its output
as
\begin{equation}
\rho_{\mathbf{ab}}^{n}=\bar{\Lambda}(\sigma^{\otimes n}), \label{stretchGEN}%
\end{equation}
for a trace-preserving LOCC $\bar{\Lambda}$.
\end{description}

By replacing Eq.~(\ref{toSImplify}) into~(\ref{stretchGEN}), and exploiting
the properties of monotonicity and subadditivity in Eqs.~(\ref{monoPROP})
and~(\ref{subPROP}), we achieve and generalize the main insight of PLOB, i.e.,
the simplification%
\begin{equation}
\mathcal{C}(\mathcal{E})\leq E^{\infty}(\sigma^{\otimes n})\leq E(\sigma),
\end{equation}
where $\sigma=\rho_{\mathcal{E}}$ if $\mathcal{E}$ is teleportation covariant.

\subsection{Channel approximations}

It is clear that the technique can be extended to bound the capacity of a
quantum channel $\mathcal{E}$ which is approximated by another channel
$\mathcal{\tilde{E}}$\ whose simulation is known and based on some resource
state $\tilde{\sigma}$. Consider two DV\ channels $\mathcal{E}$ and
$\mathcal{\tilde{E}}$ with diamond distance
\begin{equation}
||\mathcal{E}-\mathcal{\tilde{E}}||_{\diamond}\leq\delta~. \label{diamk}%
\end{equation}
For the same adaptive protocol $\mathcal{P}=\{\Lambda_{0},\ldots,\Lambda
_{n}\}$, consider the output state generated by $n$\ transmissions over these
two channels, i.e.,%
\begin{align}
\rho_{\mathbf{ab}}^{n}  &  :=\Lambda_{n}\circ\mathcal{E}\circ\Lambda
_{n-1}\cdots\circ\Lambda_{1}\circ\mathcal{E}(\rho_{\mathbf{ab}}^{0}),\\
\tilde{\rho}_{\mathbf{ab}}^{n}  &  :=\Lambda_{n}\circ\mathcal{\tilde{E}}%
\circ\Lambda_{n-1}\cdots\circ\Lambda_{1}\circ\mathcal{\tilde{E}}%
(\rho_{\mathbf{ab}}^{0}),
\end{align}
where we also have $\tilde{\rho}_{\mathbf{ab}}^{n}=\bar{\Lambda}(\tilde
{\sigma}^{\otimes n})$ by applying teleportation stretching to the simulable channel.

Using our previous \textquotedblleft peeling argument\textquotedblright, we
may evolve Eq.~(\ref{diamk}) into an error on the output state
\begin{equation}
\left\Vert \rho_{\mathbf{ab}}^{n}-\tilde{\rho}_{\mathbf{ab}}^{n}\right\Vert
\leq n\delta~.
\end{equation}
Then, assume that $\mathcal{P}$ is optimized for $\mathcal{E}$ so that
$\rho_{\mathbf{ab}}^{n}$ approximates a target state $\phi^{n}$\ with $nR_{n}$
bits, i.e., $\left\Vert \rho_{\mathbf{ab}}^{n}-\phi^{n}\right\Vert
\leq\varepsilon$. Using the triangle inequality, we write%
\begin{equation}
\left\Vert \tilde{\rho}_{\mathbf{ab}}^{n}-\phi^{n}\right\Vert \leq
\varepsilon^{\prime}:=\varepsilon+n\delta~.
\end{equation}
For small enough $\varepsilon^{\prime}$, this leads to
\begin{equation}
R_{n}\leq\frac{E(\tilde{\rho}_{\mathbf{ab}}^{n})+g(\varepsilon^{\prime}%
)\log_{2}d+h(\varepsilon^{\prime})}{n}. \label{eqlead}%
\end{equation}
Using stretching and Eq.~(\ref{sizePROP}), we have%
\begin{equation}
R_{n}\leq E(\tilde{\sigma})+\alpha_{n}g(\varepsilon^{\prime})+\frac
{h(\varepsilon^{\prime})}{n}. \label{validf}%
\end{equation}

Note that the upper bound is the same for any $\mathcal{P}$, so that it is
also true\ if we consider the supremum over $\mathcal{P}$. This is therefore
an upper bound for the $n$-use two-way capacity of the channel $\mathcal{E}$.
In other words, for $n$ uses and epsilon security $\varepsilon$, we may write
the secret key capacity
\begin{equation}
K(\mathcal{E},n,\varepsilon)\leq E(\tilde{\sigma})+\alpha_{n}g(\varepsilon
+n\delta)+\frac{h(\varepsilon+n\delta)}{n}~.
\end{equation}
It is clear that, in order to be valid, we need to have $n\delta$ small
enough. This is not a problem in the case of one-shot capacity $(n=1)$, for
which we just have
\begin{equation}
K^{(1)}(\mathcal{E},\varepsilon):=K(\mathcal{E},1,\varepsilon)\leq
E(\tilde{\sigma})+\alpha_{1}g(\varepsilon+\delta)+h(\varepsilon+\delta)~.
\end{equation}
The problem occurs for large $n$, where $n\delta$ may explode.

\subsection{Sequences of channels}

The previous problem is certainly solved in the case of a sequence of
simulable channels converging in diamond norm, i.e., for $\mathcal{\tilde{E}%
}^{\mu}$ such that%
\begin{equation}
\delta_{\mu}:=||\mathcal{E}-\mathcal{\tilde{E}}^{\mu}||_{\diamond}\overset
{\mu}{\rightarrow}0~. \label{tobbb}%
\end{equation}
In such a case we may simultaneously write%
\begin{align}
\text{Simulation error}  &  \text{:~~}\left\Vert \rho_{\mathbf{ab}}^{n}%
-\tilde{\rho}_{\mathbf{ab}}^{\mu,n}\right\Vert \leq n\delta_{\mu}\overset{\mu
}{\rightarrow}0,\label{ee1}\\
\text{Epsilon closeness}  &  \text{:~~}\left\Vert \rho_{\mathbf{ab}}^{n}%
-\phi^{n}\right\Vert \leq\varepsilon,\label{ee2}\\
\text{Stretching}  &  \text{:~~}\tilde{\rho}_{\mathbf{ab}}^{\mu,n}%
=\bar{\Lambda}(\tilde{\sigma}^{\mu\otimes n}). \label{ee3}%
\end{align}

Using the triangle inequality, we therefore have%
\begin{equation}
\left\Vert \bar{\Lambda}(\tilde{\sigma}^{\mu\otimes n})-\phi^{n}\right\Vert
\leq\varepsilon_{\mu}:=\varepsilon+n\delta_{\mu}.
\end{equation}
By applying Eqs.~(\ref{norma})-(\ref{subPROP}) and~Eq.~(\ref{sizePROP}), we
get
\begin{equation}
R_{n}\leq E(\tilde{\sigma}^{\mu})+\alpha_{n}g(\varepsilon_{\mu})+\frac
{h(\varepsilon_{\mu})}{n}.
\end{equation}
Taking the limit of large $\mu$, this becomes%
\begin{equation}
R_{n}\leq\lim_{\mu}E(\tilde{\sigma}^{\mu})+\alpha_{n}g(\varepsilon
)+\frac{h(\varepsilon)}{n}~.
\end{equation}
Then, for large $n$ and small $\varepsilon$, we find the weak converse%
\begin{equation}
\lim_{n}R_{n}\leq\lim_{\mu}E(\tilde{\sigma}^{\mu})~.
\end{equation}
By optimizing over $\mathcal{P}$, we get
\begin{equation}
K(\mathcal{E})\leq\lim_{\mu}E(\tilde{\sigma}^{\mu})~.
\end{equation}

\subsection{Infinite dimension}

The previous approach with sequences of channels is particularly useful for CV
systems. As we know from PLOB, we need to use a truncation argument which is
then released at the very end. Let us assume Alice and Bob use a
trace-preserving truncation LOCC $\mathbb{T}_{d}$ on their output state
$\rho_{\mathbf{ab}}^{n,d}=\mathbb{T}_{d}(\rho_{\mathbf{ab}}^{n})$. See
Ref.~\cite[Supplementary Note~1]{PLOB} on how to build this local CV-DV
mapping. Also assume that the input alphabet is energy-constrained, so that we
have $\mathcal{D}_{N}$ with bounded energy $N$. This latter condition may also
be realized by applying $\mathbb{T}_{d}$ before each transmission. In this
case, we will have an energy constraint depending on the truncated dimension,
i.e., $N=N(d)$.

Let us consider the energy-constrained diamond distance between the original
channel and the sequence of simulable channels $\mathcal{\tilde{E}}^{\mu}$.
For any finite $N$, \ assume that%
\begin{equation}
\delta_{\mu}^{N}:=||\mathcal{E}-\mathcal{\tilde{E}}^{\mu}||_{\diamond
N}\overset{\mu}{\rightarrow}0~.
\end{equation}
For the truncated output $\rho_{\mathbf{ab}}^{n,d}$ (approximating a target
state $\phi^{n,d}$\ with $nR_{n,d}$ bits)\ and its simulation $\tilde{\rho
}_{\mathbf{ab}}^{\mu,n,d}$ (obtained by replacing $\mathcal{E}$ with
$\mathcal{\tilde{E}}^{\mu}$\ in the protocol), we may write the following (for
any $d$ and associated $N$)%
\begin{align}
\text{Simulation error}  &  \text{:~~}\left\Vert \rho_{\mathbf{ab}}%
^{n,d}-\tilde{\rho}_{\mathbf{ab}}^{\mu,n,d}\right\Vert \leq n\delta_{\mu}%
^{N}\overset{\mu}{\rightarrow}0,\\
\text{Epsilon closeness}  &  \text{:~~}\left\Vert \rho_{\mathbf{ab}}%
^{n,d}-\phi^{n,d}\right\Vert \leq\varepsilon,\\
\text{Stretching}  &  \text{:~~}\tilde{\rho}_{\mathbf{ab}}^{\mu,n,d}%
=\bar{\Lambda}_{d}(\tilde{\sigma}^{\mu\otimes n}).
\end{align}

Using the triangle inequality, we therefore have%
\begin{equation}
\left\Vert \bar{\Lambda}_{d}(\tilde{\sigma}^{\mu\otimes n})-\phi
^{n,d}\right\Vert \leq\varepsilon_{\mu}:=\varepsilon+n\delta_{\mu}.
\end{equation}
Using Eq.~(\ref{sizePROP}) and previous reasonings, we get%
\begin{equation}
R_{n,d}\leq E(\tilde{\sigma}^{\mu})+\alpha_{n}g(\varepsilon_{\mu}%
)+\frac{h(\varepsilon_{\mu})}{n}.
\end{equation}
Taking the limit of large $\mu$, this becomes%
\begin{equation}
R_{n,d}\leq\underset{\mu}{\lim\inf}E(\tilde{\sigma}^{\mu})+\alpha
_{n}g(\varepsilon)+\frac{h(\varepsilon)}{n}~,
\end{equation}
where we use the inferior limit to account for the fact that $\tilde{\sigma
}^{\mu}$ may be an unbounded sequence of states.

Then, for large $n$ and small $\varepsilon$, we find%
\begin{equation}
\lim_{n}R_{n,d}\leq\underset{\mu}{\lim\inf}E(\tilde{\sigma}^{\mu})~.
\end{equation}
By optimizing over the protocols $\mathcal{P}$ (i.e., over the equivalent
exponential-size class of $\mathcal{P}$), we then get
\begin{equation}
K(\mathcal{E}|N):=\sup_{\mathcal{P}}\lim_{n}R_{n,d}\leq\underset{\mu}{\lim
\inf}E(\tilde{\sigma}^{\mu})~.
\end{equation}
It is clear that the upper bound does not depend on the constraint $N$, so
that this constraint (and the truncation) can be relaxed. In other words, we
have%
\begin{equation}
K(\mathcal{E}):=\sup_{N}K(\mathcal{E}|N)\leq\underset{\mu}{\lim\inf}%
E(\tilde{\sigma}^{\mu})~. \label{eqmm}%
\end{equation}

Note that this result holds for an entanglement measure $E$ with the desired
properties above, such as the squashed entanglement or the REE. If we specify
the result to the REE, then this procedure is an alternate proof of the one
given in Sec.~\ref{SEC_CV_asy}. Note that, setting $E=E_{R}$, Eq.~(\ref{eqmm})
becomes
\begin{align}
K(\mathcal{E})  &  \leq\underset{\mu}{\lim\inf}E_{R}(\tilde{\sigma}^{\mu
})=\underset{\mu}{\lim\inf}\inf_{\sigma_{s}}S(\tilde{\sigma}^{\mu}||\sigma
_{s})\nonumber\\
&  =\inf_{\sigma_{s}^{\mu}}\underset{\mu}{\lim\inf}S(\tilde{\sigma}^{\mu
}||\sigma_{s}^{\mu}):=E_{R}(\tilde{\sigma}),
\end{align}
where we use the definition in Eq.~(\ref{REE_weaker}) with $\tilde{\sigma
}:=\lim_{\mu}\tilde{\sigma}^{\mu}$.

\section{Literature on channel simulation and protocol reduction}

Let us here discuss the precursory ideas that were in the literature before
the full generalization devised in PLOB. Besides this section, one may also
read the Supplementary Notes 8 and 9 in Ref.~\cite{PLOB}. A summary of the
following discussion is given in Table~\ref{TABLEcomp}, where we make a direct
comparison between PLOB and previous approaches and methodologies.

\begin{table*}[ptb]
\centering%
\[%
\begin{tabular}
[c]{c|ccccc}
& $%
\begin{array}
[c]{c}%
\text{BDSW, HHH99}\\
\text{\cite{B2main,HoroTEL}}%
\end{array}
$ & $%
\begin{array}
[c]{c}%
\text{MH12,W12}\\
\text{\cite{MHthesis,Wolfnotes}}%
\end{array}
$ & $%
\begin{array}
[c]{c}%
\text{LM15}\\
\text{\cite{Leung}}%
\end{array}
$ & $%
\begin{array}
[c]{c}%
\text{GC02,NFC09}\\
\text{\cite{Cirac,Niset}}%
\end{array}
$ & $%
\begin{array}
[c]{c}%
\text{PLOB}\\
\text{\cite{PLOB}}%
\end{array}
$\\\hline
&  &  &  &  & \\
$%
\begin{array}
[c]{c}%
\text{Simulated}\\
\text{channels}%
\end{array}
$ & $%
\begin{array}
[c]{c}%
\text{Pauli}\\
\text{channels }\mathcal{P}\text{ \cite{B2main}.}\\
\text{Sub-class of}\\
\text{DV channels \cite{HoroTEL}}%
\end{array}
$ & $%
\begin{array}
[c]{c}%
\text{All DV\ channels}\\
\text{but probabilistically.}\\
\text{If tele-covariant, then}\\
\text{deterministically}%
\end{array}
$ & $%
\begin{array}
[c]{c}%
\text{Tele-covariant}\\
\text{DV\ channels}%
\end{array}
$ & $%
\begin{array}
[c]{c}%
\text{Gaussian}\\
\text{channels}%
\end{array}
$ & $%
\begin{array}
[c]{c}%
\text{Any channel}\\
\text{(DV \& CV)}\\
\text{LOCC simulable}\\
\text{by resource state $\sigma$}%
\end{array}
$\\
&  &  &  &  & \\
$%
\begin{array}
[c]{c}%
\text{Amplitude}\\
\text{damping}%
\end{array}
$ & $%
\begin{array}
[c]{c}%
\text{Not}\\
\text{simulable}%
\end{array}
$ & $%
\begin{array}
[c]{c}%
\text{Probabilistically}\\
\text{simulable}%
\end{array}
$ & $%
\begin{array}
[c]{c}%
\text{Not}\\
\text{simulable}%
\end{array}
$ & $%
\begin{array}
[c]{c}%
\text{Not}\\
\text{simulable}%
\end{array}
$ & Simulable\\
&  &  &  &  & \\
Criterion & N/A & $%
\begin{array}
[c]{c}%
\text{Tele-covariance}\\
\text{(for DV)}%
\end{array}
$ & $%
\begin{array}
[c]{c}%
\text{Tele-covariance}\\
\text{(for DV)}%
\end{array}
$ & N/A & $%
\begin{array}
[c]{c}%
\text{Tele-covariance}\\
\text{(for DV \& CV)}%
\end{array}
$\\
&  &  &  &  & \\
$%
\begin{array}
[c]{c}%
\text{Simulation}\\
\text{error}%
\end{array}
$ & N/A & $%
\begin{array}
[c]{c}%
\text{Probability of }\\
\text{teleportation}%
\end{array}
$ & N/A & $%
\begin{array}
[c]{c}%
\text{Not}\\
\text{controlled}%
\end{array}
$ & $%
\begin{array}
[c]{c}%
\text{Yes. Controlled}\\
\text{for CV channels}%
\end{array}
$\\
&  &  &  &  & \\
$%
\begin{array}
[c]{c}%
\text{Protocol}\\
\text{task}%
\end{array}
$ & QC & QC & QC & QC & $%
\begin{array}
[c]{c}%
\text{Any task}\\
\text{(QC, ED, QKD)}%
\end{array}
$\\
&  &  &  &  & \\
$%
\begin{array}
[c]{c}%
\text{Type of}\\
\text{reduction}%
\end{array}
$ & $%
\begin{array}
[c]{c}%
\text{QC}\rightarrow\text{ED}\\
\text{Reduction to }\\
\text{ent. distillation}%
\end{array}
$ & $%
\begin{array}
[c]{c}%
\text{QC}\rightarrow\text{ED}\\
\text{Reduction to }\\
\text{ent. distillation}%
\end{array}
$ & $%
\begin{array}
[c]{c}%
\text{QC}\rightarrow\text{PPT}\\
\text{Reduction to }\\
\text{PPT distillation}%
\end{array}
$ & $%
\begin{array}
[c]{c}%
\text{QC}\rightarrow\text{ED}\\
\text{with Gaussian}\\
\text{LOCCs~\cite{Niset}}%
\end{array}
$ & $%
\begin{array}
[c]{c}%
\text{Adaptive protocol}\\
\rightarrow\text{block protocol.}\\
\text{Task-preserving}\\
\text{(any dim, DV/CV)}%
\end{array}
$\\
&  &  &  &  & \\
$%
\begin{array}
[c]{c}%
\text{Type of }\\
\text{bound}%
\end{array}
$ & $%
\begin{array}
[c]{c}%
Q_{1}(\mathcal{P})\leq D_{1}(\rho_{\mathcal{P}})\\
\text{(extended to 2-way}\\
\text{CCs) \cite[Sec. V]{B2main}}%
\end{array}
$ & $%
\begin{array}
[c]{c}%
Q_{2}(\mathcal{E})\leq d^{2}D_{2}(\rho_{\mathcal{E}})\\
\text{in finite dim }d\\
\text{\cite[Theorem~14]{MHthesis}}%
\end{array}
$ & $%
\begin{array}
[c]{c}%
\text{Bounds to DV\ quantum}\\
\text{capacities restricted to }\\
\text{PPT-preserving codes}%
\end{array}
$ & N/A & $%
\begin{array}
[c]{c}%
Q_{2}(\mathcal{E})\leq K(\mathcal{E})\leq E_{R}(\sigma)\\
\sigma=\rho_{\mathcal{E}}\text{ if tele-covariant}\\
\text{(any dim, DV/CV)}%
\end{array}
$%
\end{tabular}
\ \ \
\]
\caption{Comparison between PLOB and previous literature on channel simulation
and protocol reduction.}%
\label{TABLEcomp}%
\end{table*}

The first insight was introduced in 1996 by BDSW~\cite{B2main}. This was based
on the standard teleportation protocol for DV\ systems and allowed these
authors to simulate \textquotedblleft generalized depolarizing
channels\textquotedblright, later known as Pauli channels~\cite{NielsenChuang}%
. The restriction of this original technique to Pauli channels was first shown
in Ref.~\cite{SougatoBowen} and later re-examined in Ref.~\cite{Tompaper}.
BDSW first recognized that a Pauli channel $\mathcal{P}$ can be simulated by
teleporting over its Choi matrix $\rho_{\mathcal{P}}$. See
Ref.~\cite[Section~V]{B2main}. This specific case was later re-considered as a
property of mutual \textit{reproducibility }between states and
channels~\cite{HoroTEL}. Let us remark that Ref.~\cite{HoroTEL} also explored
the possibility to extend channel simulation beyond teleportation by using
more general LOCCs. In principle, this allowed them to simulate more channels
but still a sub-class of DV channels, due to the specific use of
finite-dimensional and non-asymptotic LOCCs (e.g., see Eq.~(11) in
Ref.~\cite{HoroTEL}).

Similar simulation ideas, but in the setting of quantum computing, were
considered in Ref.~\cite{Gottesman} (see also the more recent
Ref.~\cite{Aliferis})\ where a unitary $U$ is stored in its Choi matrix
$\rho_{U}$. This unitary is then applied to some input state $\rho$ by
teleporting such input over $\rho_{U}$. This is also known as
\textquotedblleft quantum gate teleportation\textquotedblright. It shows that
teleportation is a primitive for quantum computation. Likewise, teleportation
can be expressed in terms of primitive quantum computational
operations~\cite{SamBrassard}. Quantum gate teleportation is also at the heart
of linear-optical quantum computing based on linear optics and probabilistic
gates~\cite{Knill} (see also Ref.~\cite{telereview} for a general overview on
these applications of teleportation).

Using the teleportation simulation of a Pauli channel $\mathcal{P}$, BDSW
first showed how to transform a quantum communication (QC) protocol into an
entanglement distillation (ED) protocol over its Choi matrix $\rho
_{\mathcal{P}}$. We call this technique \textquotedblleft reduction to
entanglement distillation\textquotedblright. This allowed them to prove the
following bound on the one-way quantum capacity
\begin{equation}
Q_{1}(\mathcal{P})\leq D_{1}(\rho_{\mathcal{P}}), \label{ub1BENNETT}%
\end{equation}
where $D_{1}$ is the one-way distillability. This result was implicitly
extended to two-way CC, so that they also wrote
\begin{equation}
Q_{2}(\mathcal{P})\leq D_{2}(\rho_{\mathcal{P}}). \label{ubBENNETT}%
\end{equation}
Reduction to entanglement distillation (QC$\rightarrow$ED) was originally
formulated in an asymptotic fashion, i.e., for large $n$, which is sufficient
to prove Eqs.~(\ref{ub1BENNETT}) and~(\ref{ubBENNETT}).

More recently, in 2012, Refs.~\cite{MHthesis,Wolfnotes} considered the\textit{
probabilistic} simulation of an arbitrary DV quantum channel $\mathcal{E}$ via
teleportation. This is done by assuming that Alice and Bob only picks the Bell
outcome corresponding to the identity operator, which occurs with probability
$p=d^{-2}$, where $d$ is the dimension of the input system. This version can
also be traced back to the probabilistic approach of Ref.~\cite{Knill}. In the
presence of a probability of success associated with the simulation, one can
derive upper bounds similar to those of BDSW but with a suitable pre-factor.
In fact, adopting the probabilistic simulation and BDSW's reduction to
entanglement distillation (QC$\rightarrow$ED), Ref.~\cite{MHthesis} showed
\begin{equation}
Q_{2}(\mathcal{E})\leq p^{-1}D_{2}(\rho_{\mathcal{E}}), \label{MH}%
\end{equation}
for an arbitrary DV channel $\mathcal{E}$. Let us remark that
Refs.~\cite{MHthesis,Wolfnotes} also identified the property of teleportation
covariance for DV\ channels, realizing that these channels can be simulated
deterministically, i.e., with an associated success probability $p=1$.

In 2015, Ref.~\cite{Leung} too identified the criterion of teleportation
covariance of DV channels and considered the (deterministic) simulation of
such channels over their Choi matrices. In particular, Ref.~\cite{Leung}
assumed the possibility of more general teleportation protocols as those
introduced in Ref.~\cite{WernerTELE}. Because these simulations are
non-asymptotic, the class of DV channels is limited and, for instance, it
cannot include the amplitude damping channel. Ref.~\cite{Leung} adopted a
variation of the BDSW argument to simplify quantum communication. In fact,
they showed how to simplify positive-partial transpose (PPT) preserving codes
over a teleportation covariant channel into PPT-distillation over copies of
its Choi matrix. Thanks to this \textquotedblleft reduction to PPT
distillation\textquotedblright, they were able to write one-shot upper bounds
for PPT-preserving code quantum capacities.

In the framework of CV systems, Ref.~\cite{Cirac} first studied the simulation
of single-mode Gaussian channels by using the BK protocol. Due to the nature
of the topics studied in that paper (which is about the impossibility to
distill entanglement from Gaussian entangled states with Gaussian LOCCs), no
control of the simulation error was considered. The same approach was later
followed by Ref.~\cite{Niset}. The latter used the channel simulation to
reduce a Gaussian quantum error correcting code into Gaussian entanglement distillation.

Within this general context, PLOB introduced the most general type of channel
simulation in a quantum communication scenario, where an LOCC and a resource
state are used to simulate an arbitrary quantum channel at any dimension
(finite or infinite). See Eqs.~(\ref{sigma00})-(\ref{asy2}). PLOB\ also
established teleportation covariance as a criterion to identify
Choi-stretchable (i.e., teleportation-simulable) channels at any dimension. In
particular, PLOB extended the technique by developing a rigorous theory of
asymptotic channel simulation, which is crucial not only for bosonic channels
but also for the deterministic asymptotic simulation of DV\ channels, such as
the amplitude damping channel.

Using channel simulation, PLOB showed how to simplify an arbitrary adaptive
protocol implemented over an arbitrary channel at any dimension, finite or
infinite (teleportation stretching).\ Differently from previous approaches
(which were about reduction to entanglement distillation), teleportation
stretching works by preserving the original communication task. This means
that\ an adaptive protocol of quantum communication (QC), entanglement
distribution (ED) or quantum key distribution (QKD), is reduced to a
corresponding block protocol with exactly the same original task (QC, ED, or
QKD). In particular, the output state is decomposed in terms of a tensor
product of resource states as in Eqs.~(\ref{StretchingMAIN})
and~(\ref{stretchASYm}).

The adaptive-to-block reduction of a private communication protocol has been
first introduced in PLOB. Most importantly, PLOB has shown how to combine this
reduction with the properties of an entanglement measure as the REE. The
entire recipe of \textquotedblleft REE plus teleportation
stretching\textquotedblright\ has led to the determination of the tightest
known upper bound for the secret key capacity (and the other two-way assisted
capacities) of a quantum channel at any dimension. See Eqs.~(\ref{UB1})
and~(\ref{UBcvm}).

These techniques developed by PLOB were picked up and exploited in a series of
follow-up papers, including WTB~\cite{WildeFollowup}. More recently,
Ref.~\cite{GerLimited} introduced a simulation of Gaussian channels based on
finite-energy resource states. This was promptly combined with the techniques
of PLOB in Ref.~\cite{RicFINITE}\ to derive sub-optimal approximations of
previously-established weak converse bounds for private communication. Finally
note that the non-local simulations~\cite{Qsim0,Qsim1,Qsim2} based on
deterministic versions of the programmable quantum gate array~\cite{Gatearray}
are clearly not suitable for quantum and private communication where Alice and
Bob can only implement LOCCs.

\section{Strong converse rates}

\subsection{Preliminary comments}

At the end of February 2016, four months after the first version of PLOB
appeared on the arXiv, the follow-up paper WTB~\cite{WildeFollowup} also
appeared. An explicit timeline of the contributions is provided in
Table~\ref{TABLEline}\ for the sake of clarity. As we can see, the first
version of PLOB~\cite{PLOB} appeared in October 2015. The first arXiv version
of PLOB already contained the most important result for the pure-loss channel
(PLOB bound). Full details of the methodology were included in the second
arXiv version in December 2015~\cite{PLOBv2}. All the other two-way capacities
and bounds were collected in a twin paper~\cite{PL} which also appeared in
December 2015 and was later merged in the published version of
PLOB~\cite{PLOB}. In a few words, the main results were all proven in 2015,
well before the appearance of WTB. Subsequent arXiv versions of PLOB only
added refinements and minor clarifications.

\begin{table*}[ptb]
\centering%
\[%
\begin{tabular}
[c]{llll}%
Date: & ~ & Manuscripts on the arXiv: & Main contents:\\\hline
~ & ~ &  & \\
29 Oct 2015 &  & $\text{First version of PLOB~\cite{PLOB}}$ & $%
\begin{tabular}
[c]{l}%
$\text{Introduces }$the secret-key capacity of the lossy channel\\
(PLOB bound) $-\log_{2}(1-\eta)$.
\end{tabular}
$\\
~ &  &  & \\
8 Dec 2015 &  & Second$\text{ version of PLOB~\cite{PLOBv2}}$ & $%
\begin{tabular}
[c]{l}%
Includes$\text{ the general methodology: (i) t}$he REE weak converse bound\\
and (ii) its reduction by teleportation stretching to single letter.\\
PLOB bound extended to the thermal-loss channel.
\end{tabular}
$\\
~ &  &  & \\
15 Dec 2015 &  & $%
\begin{tabular}
[c]{l}%
First$\text{ version of Ref.~\cite{PL}}$\\
(merged in published PLOB)
\end{tabular}
$ & $%
\begin{tabular}
[c]{l}%
Extends the results to all teleportation-covariant channels, including:\\
Pauli, erasure channels, and bosonic Gaussian channels.
\end{tabular}
$\\
~ &  &  & \\
5 Jan 2016 &  & $%
\begin{tabular}
[c]{l}%
Third version of PLOB\\
and first version of$\text{ Ref.~\cite{ref1}}$%
\end{tabular}
\ $ & $%
\begin{tabular}
[c]{l}%
Ref.$~\text{\cite{ref1}}$ extends methods and results of PLOB to
repeater-assisted\\
quantum communications and arbitrary quantum networks.
\end{tabular}
$\\
~ &  &  & \\
29 Feb 2016 &  & \underline{First version} of$\text{ WTB~\cite{WildeFollowup}%
}$ & $%
\begin{tabular}
[c]{l}%
Use methods of PLOB to study the strong converse property of the\\
bounds established in PLOB for teleportation-covariant channels.
\end{tabular}
$%
\end{tabular}
\ \ \
\]
\caption{Timeline of the main results established in the early arXiv versions
of PLOB, before the appearance of the follow-up analysis by WTB on the arXiv.}%
\label{TABLEline}%
\end{table*}

Using the methodology devised in PLOB, WTB studied how the
previously-established weak converse bounds for teleportation-covariant
channels are also strong converse bounds. In private communication, a weak
converse bound means that \textit{perfect} secret keys cannot be established
at rates above the bound. A strong converse bound is a refinement according to
which even \textit{imperfect} secret keys ($\varepsilon$-secure with
$\varepsilon>0$) cannot be shared at rates above the bound for many uses. Let
us clarify some important points about this paper besides discussing and
fixing its technical error.

Even though WTB does not adopt the terminology introduced by PLOB
(teleportation stretching, stretchable channels etc.), one can easily check
that WTB exploits exactly the methodology previously introduced by PLOB. In
fact, WTB combines the following ingredients

\begin{itemize}
\item A notion of channel's REE to bound key generation

\item Teleportation stretching to simplify adaptive protocols for private communication.
\end{itemize}

\noindent In a few words, WTB adopts the entire reduction idea of PLOB, which
is based on using channel's REE on top of teleportation stretching. This is
what allows them to write single-letter upper bounds.

To be more precise, WTB first defines \textquotedblleft classical pre- and
post-processing (CPPP) protocols\textquotedblright. These are \textit{non}%
-adaptive protocols where the remote parties are limited to a single rounds of
initial and final LOCCs. In this context, they derive strong converse rates
for CPPP-assisted private communication (see Ref.~\cite[Theorem~13]%
{WildeFollowup}). To generalize the approach and include adaptive protocols
with unlimited two-way CCs (over teleportation-covariant channels), they then
employ channel's REE and teleportation stretching. This allows them to write
their Theorems~12 and~19, which are the strong converse versions of
Ref.~\cite[Theorem~5]{PLOB} in PLOB.

Indeed, for a teleportation-covariant channel $\mathcal{E}$, WTB wrote the
strong converse bound~\cite[Theorem~19]{WildeFollowup}
\begin{equation}
K(\mathcal{E})\leq\Phi(\mathcal{E})+\sqrt{\frac{V(\mathcal{E})}{n}}%
\varphi^{-1}(\varepsilon)+\mathcal{O}\left(
\frac{\log_2{n}}{n}\right)  ,
\label{fffjjjj}%
\end{equation}
for $n\geq1$ channel uses and security parameter $\varepsilon\in(0,1)$, where
$\Phi(\mathcal{E})=E_{R}(\rho_{\mathcal{E}})$ is the weak converse bound
established in PLOB, and
\begin{equation}
\varphi(a)=\int_{-\infty}^{a}dx\,e^{-x^{2}/2}/\sqrt{2\pi}.
\end{equation}
The entropic variance $V(\mathcal{E})$ in Eq.~(\ref{fffjjjj}) is defined as%
\begin{equation}
V(\mathcal{E})=\left\{
\begin{array}
[c]{c}%
\sup_{\sigma_{s}}V(\rho_{\mathcal{E}}||\sigma_{s}),~~\text{for~}%
2\varepsilon<1,\\
\\
\inf_{\sigma_{s}}V(\rho_{\mathcal{E}}||\sigma_{s}),~~\text{for~}%
2\varepsilon\geq1,
\end{array}
\right.
\end{equation}
where $V(\rho||\sigma)=\mathrm{Tr}\left\{
\rho\lbrack\log_2\rho-\log_2 \sigma-S(\rho||\sigma)]^{2}\right\}
$, and the supremum/infimum are taken over the set of separable
states $\sigma_{s}$ that achieve the minimum in
$E_{R}(\rho_{\mathcal{E}})$.

Using a Chebyshev-like bound, one may write the strong converse bound also
as~\cite{WildeFollowup}
\begin{equation}
K(\mathcal{E})\leq\Phi(\mathcal{E})+\sqrt{\frac{V(\mathcal{E})}%
{n(1-\varepsilon)}}+\frac{C(\varepsilon)}{n}~, \label{strongconv}%
\end{equation}
where
\begin{equation}
C(\varepsilon):=\log_{2}6+2\log_{2}\left(  \frac{1+\varepsilon}{1-\varepsilon
}\right)  .
\end{equation}
As a consequence, the weak-converse bounds for dephasing, erasure and other DV
channels established in PLOB would also be strong converse rates, according to
Propositions~22 and 23 stated in WTB~\cite{WildeFollowup}.

Finally, WTB~\cite{WildeFollowup} attempts to generalize the above result to
CV systems in Theorem~24. If their Theorem~24 were true, then also the PLOB
bounds for Gaussian channels would be strong converse rates. Unfortunately,
WTB does \textit{not} rigorously prove its Theorem~24. In fact, following an
incorrect interpretation of the BK protocol, WTB assumes that CV teleportation
asymptotically induces a perfect quantum channel (i.e., an identity channel)
independently from the size of the input alphabet of quantum states. By
contrast, we know that the BK teleportation channel \textit{does not}
uniformly converge to the identity channel. As a result, the bounds for
Gaussian channels stated in Ref.~\cite[Theorem~24]{WildeFollowup} are
technically equal to infinity.

\subsection{Claims and mathematical issues}

Let us describe the problem in detail.
WTB\ makes the following equivalent claims on the strong-converse bound.

\begin{itemize}
\item \textbf{WTB claim~(\cite[Theorem~24]{WildeFollowup}).}~Consider an
$\varepsilon$-secure key generation protocol over $n$ uses of a
phase-insensitive Gaussian channel $\mathcal{E}$, which may be a thermal-loss
channel ($\mathcal{E}_{\eta,\bar{n}}$), a quantum amplifier ($\mathcal{E}%
_{g,\bar{n}}$) or an additive-noise Gaussian channel ($\mathcal{E}_{\xi}$).
For any $\varepsilon\in(0,1)$ and $n\geq1$, one has the upper bound of
Eq.~(\ref{strongconv}) for the secret key rate, where $\Phi(\mathcal{E})$ is
the weak-converse bound established in PLOB, and the \textquotedblleft
unconstrained relative entropy variance\textquotedblright\ $V(\mathcal{E})$ is
respectively given by
\begin{align}
V(\mathcal{E}_{\eta,\bar{n}})  &  =\bar{n}(\bar{n}+1)\log_{2}^{2}\left[
\eta(\bar{n}+1)/\bar{n}\right]  ,\nonumber\\
V(\mathcal{E}_{g,\bar{n}})  &  =\bar{n}(\bar{n}+1)\log_{2}^{2}\left[
g^{-1}(\bar{n}+1)/\bar{n}\right] \nonumber\\
V(\mathcal{E}_{\xi})  &  =(1-\xi)^{2}/\ln^{2}2~.
\end{align}
In particular, for a pure loss channel ($\mathcal{E}_{\eta,0}$) and a
quantum-limited amplifier ($\mathcal{E}_{g,0}$), one has%
\begin{equation}
K(\mathcal{E})\leq\Phi(\mathcal{E})+\frac{C(\varepsilon)}{n}~. \label{cc11}%
\end{equation}

\item \textbf{WTB\ claim~(\cite{WildeFollowup}, with simulation error).}~The
above claim is obtained starting from a finite simulation energy $\mu$ and
then taking the limit of $\mu\rightarrow\infty$. For any security parameter
$\varepsilon\in(0,1)$, number of channel uses $n\geq1$ and simulation energy
$\mu$ with \textquotedblleft infidelity\textquotedblright\ $\varepsilon
_{\text{TP}}(n,\mu)$, one may write the following upper bound for the secret
key rate of a phase insensitive Gaussian channel $\mathcal{E}$%
\begin{equation}
K(\mathcal{E})\leq\Phi(\mathcal{E})+\Delta(n,\mu), \label{followPLOB}%
\end{equation}
where $\Phi(\mathcal{E})$ is the weak-converse bound established in PLOB. Here
$\Delta(n,\mu)$ has the asymptotic expansion%
\begin{align}
\Delta(n,\mu)  &  \simeq\sqrt{\frac{2V(\mathcal{E})+O(\mu^{-1})}%
{n[1-\varepsilon(n,\mu)]}}\nonumber\\
&  +\frac{C[\varepsilon(n,\mu)]}{n}+O(\mu^{-1})~, \label{eqrr}%
\end{align}
at fixed $n$ and large $\mu$, where%
\begin{equation}
\varepsilon(n,\mu):=\min\left\{  1,\left[  \sqrt{\varepsilon}+\sqrt
{\varepsilon_{\text{TP}}(n,\mu)}\right]  ^{2}\right\}  . \label{errorTT}%
\end{equation}
For a pure loss channel ($\mathcal{E}_{\eta,0}$) and a quantum-limited
amplifier ($\mathcal{E}_{g,0}$), one has Eq.~(\ref{followPLOB}), with
\begin{equation}
\Delta(n,\mu)\simeq n^{-1}C[\varepsilon(n,\mu)]+O(\mu^{-1}). \label{eqrr2}%
\end{equation}

\end{itemize}

\noindent In the previous claim, the problem is the infidelity\ parameter%
\begin{equation}
\varepsilon_{\text{TP}}(n,\mu):=1-F(\rho_{\mathbf{ab}}^{n},\rho_{\mathbf{ab}%
}^{\mu,n}), \label{pb1}%
\end{equation}
between the output of the protocol $\rho_{\mathbf{ab}}^{n}$\ and the output of
the simulated protocol $\rho_{\mathbf{ab}}^{\mu,n}$ [in WTB denoted with
$\zeta_{AB}^{n}$ and $\zeta_{AB}^{\prime}(n,\mu)$]. In fact, WTB (wrongly)
argues that~\cite{WildeFollowup}%
\begin{equation}%
\begin{array}
[c]{c}%
\text{\textquotedblleft continuous variable\ teleportation}\\
\text{induces a perfect quantum channel}\\
\text{when infinite energy is available,\textquotedblright}%
\end{array}
\label{stateWTB}%
\end{equation}
and then writes~\cite[Eq.~(178)]{WildeFollowup}%
\begin{equation}
\underset{\mu}{\lim\sup}~\varepsilon_{\text{TP}}(n,\mu)=0,~~\text{for any
}n\text{.} \label{pb2}%
\end{equation}
The fact that this quantity goes to zero is a crucial step in WTB's proof. In
fact, if this is true, then we may write $\lim_{\mu}\varepsilon(n,\mu
)=\varepsilon$ and safely replace this in Eq.~(\ref{eqrr}). By contrast, if
Eq.~(\ref{pb2}) does not hold and we get
\begin{equation}
\underset{\mu}{\lim\sup}~\varepsilon_{\text{TP}}(n,\mu)=1,~~\text{for any
}n\text{,} \label{pb2WTB}%
\end{equation}
then $\lim_{\mu}\varepsilon(n,\mu)=1$, and we have $\Delta(n,\mu)=\infty$ both
in Eqs.~(\ref{eqrr}) and~(\ref{eqrr2}). In this case, one would have proven
the trivial upper bound%
\begin{equation}
K(\mathcal{E})\leq\Phi(\mathcal{E})+\infty. \label{WTBresult}%
\end{equation}
Unfortunately, Eq.~(\ref{pb2WTB}) is the actual technical result which can be
derived following the steps of the proof presented in WTB~\cite{WildeFollowup}%
. This means that WTB proves the trivial bound in Eq.~(\ref{WTBresult}), not
Eq.~(\ref{strongconv}) or Eq.~(\ref{cc11}).

\subsection{Technical gap and exploding bound}

The first problem is that Eq.~(\ref{pb2}) is essentially given without any
mathematical derivation. To be more precise, it is not proven how the error
\textquotedblleft$\mathcal{E}^{\mu}\neq\mathcal{E}$\textquotedblright\ in the
simulation of the Gaussian channel $\mathcal{E}$ (in each single
transmission)\ is propagated into an overall error \textquotedblleft%
$\rho_{\mathbf{ab}}^{\mu,n}\neq\rho_{\mathbf{ab}}^{n}$\textquotedblright\ for
the $n$-use output of the adaptive protocol $\rho_{\mathbf{ab}}^{n}$, which is
exactly what $\varepsilon_{\text{TP}}$\ is about according to the definition
in Eq.~(\ref{pb1}). For instance, what is the dependence of such an output
error with respect to the number $n$ of channel uses? Can this error explode?

Said in other words, the fundamental gap in WTB's proof is the absence of a
peeling argument~\cite{PLOB} as the one discussed in Sec.~\ref{subsecSTRE}
[See Eq.~(\ref{casen2})], which shows how the simulation error on the channel
$\Vert\mathcal{E}-\mathcal{E}^{\mu}\Vert_{\Diamond N}$ propagates through the
adaptive protocol and is transformed into a corresponding simulation error on
the $n$-use output state $\Vert\rho_{\mathbf{ab}}^{n}-\rho_{\mathbf{ab}%
}^{n,\mu}\Vert$. This is a crucial technique for the simplification of an
adaptive protocol~\cite{PLOB}, which is based on a suitable combination of
triangle inequality and data processing (monotonicity) of the relative entropy.

Because of the absence of any peeling argument able to quantify the infidelity
$\varepsilon_{\text{TP}}(n,\mu)$ in terms of the channel simulation error
\textquotedblleft$\mathcal{E}^{\mu}\neq\mathcal{E}$\textquotedblright, we may
safely say that WTB's proof does not really apply to adaptive protocols. As
further discussed in Ref.~\cite{PLB}, peeling arguments could be formulated
assuming various topologies of convergences (strong, uniform, or
bounded-uniform) but none of these formulations can be found in WTB, where the
peculiar convergence properties of the BK\ protocol are clearly not known and
completely ignored.

The second problem is that Eq.~(\ref{pb2}) is not even proven for a single use
($n=1$), and the reasoning followed in WTB leads exactly to the opposite
result of Eq.~(\ref{pb2WTB}). In fact, let us assume a single use ($n=1$) of a
trivial adaptive protocol ($\Lambda_{1}=\mathcal{I}$), so that%
\begin{equation}
\rho_{\mathbf{ab}}^{1}=\mathcal{E}(\rho_{\mathbf{ab}}^{0}),~\rho_{\mathbf{ab}%
}^{1,\mu}=\mathcal{E}^{\mu}(\rho_{\mathbf{ab}}^{0}),
\end{equation}
where $\rho_{\mathbf{ab}}^{0}$ is the initial state of the
registers, and the channels are meant to be applied to the input
system $a_{1}$, i.e.,
$\mathcal{E}=\mathcal{I}_{\mathbf{a}}\otimes\mathcal{E}_{a_{1}}\otimes
\mathcal{I}_{\mathbf{b}}$ and $\mathcal{E}^{\mu}=\mathcal{I}_{\mathbf{a}%
}\otimes\mathcal{E}_{a_{1}}^{\mu}\otimes\mathcal{I}_{\mathbf{b}}$. Then, we
may write their infidelity as%
\begin{align}
\varepsilon_{\text{TP}}(1,\mu)  &  =1-F(\rho_{\mathbf{ab}}^{1},\rho
_{\mathbf{ab}}^{\mu,1})\\
&  =1-F[\mathcal{E}(\rho_{\mathbf{ab}}^{0}),\mathcal{E}^{\mu}(\rho
_{\mathbf{ab}}^{0})]\\
&  \geq1-F[\rho_{\mathbf{ab}}^{0},\mathcal{I}_{a_{1}}^{\mu}(\rho_{\mathbf{ab}%
}^{0})], \label{eqMMM}%
\end{align}
where we exploit the monotonicity of the fidelity under the maps
$\mathcal{E}=\mathcal{E}\circ\mathcal{I}$ and $\mathcal{E}^{\mu}%
=\mathcal{E}\circ\mathcal{I}^{\mu}$.

Now the \textquotedblleft proof idea\textquotedblright\ in WTB is based on the
statement in~(\ref{stateWTB}), which is unfortunately not sufficient to send
$\varepsilon_{\text{TP}}(1,\mu)$ to zero in the limit of large $\mu$. In fact,
this statement fails if, at the input, we consider asymptotic states whose
energy $\tilde{\mu}$ \textquotedblleft competes\textquotedblright\ with the
one $\mu$ of the resource state. It is clear that these states need to be
included among all possible inputs, because we are studying
\textit{unconstrained} quantum and private capacities, for which the input
alphabet is energy-unbounded.

Therefore, as possible input, assume that Alice sends part $a_{1}$ of a TMSV
state $\Phi_{aa_{1}}^{\tilde{\mu}}$ with energy $\tilde{\mu}$. This means that
we may decompose $\rho_{\mathbf{ab}}^{0}=\rho_{\mathbf{a}}^{0}\otimes
\Phi_{aa_{1}}^{\tilde{\mu}}\otimes\rho_{\mathbf{b}}^{0}$, and write%
\begin{equation}
\varepsilon_{\text{TP}}(1,\mu)\geq1-F[\Phi_{aa_{1}}^{\tilde{\mu}}%
,\mathcal{I}_{a}\otimes\mathcal{I}_{a_{1}}^{\mu}(\Phi_{aa_{1}}^{\tilde{\mu}%
})], \label{firstCC}%
\end{equation}
by using the multiplicativity of the fidelity. Now, from Secs.~\ref{SECbos111}
and~\ref{secGAUSS}, we know that, depending on the order of the limits, we may
write the two opposite results
\begin{equation}
\lim_{\tilde{\mu}}\lim_{\mu}\varepsilon_{\text{TP}}(1,\mu)\geq0
\end{equation}
and%
\begin{equation}
\lim_{\mu}\lim_{\tilde{\mu}}\varepsilon_{\text{TP}}(1,\mu)=1.
\end{equation}

In WTB there is no consideration of the unboundedness of the input alphabet,
and the authors just consider $\lim\sup_{\mu}$~$\varepsilon_{\text{TP}}$. This
generic limit does not imply any specific order of the limits between the
simulation energy $\mu$ and the input energy $\tilde{\mu}$ of the alphabet.
Therefore, for an unbounded alphabet, one must have%
\begin{equation}
\underset{\mu}{\lim\sup~}\varepsilon_{\text{TP}}=\max\{\lim_{\mu}\lim
_{\tilde{\mu}}\varepsilon_{\text{TP}},\lim_{\tilde{\mu}}\lim_{\mu}%
\varepsilon_{\text{TP}}\}~. \label{WTBambiguity}%
\end{equation}
It is clear that this leads to%
\begin{equation}
\underset{\mu}{\lim\sup~}\varepsilon_{\text{TP}}(1,\mu)=1~.
\end{equation}
By extending the reasoning to arbitrary $n$ (via the peeling argument), one
obtains Eq.~(\ref{pb2WTB}) and, therefore, the explosion of the bound as in
Eq.~(\ref{WTBresult}).

Here it is important to remark that the ambiguity in Eq.~(\ref{WTBambiguity})
is not addressed or noted in any part of WTB. In WTB there is no discussion
related to uniform convergence, associated with the first order of the limits
in Eq.~(\ref{WTBambiguity}), or strong convergence, associated with the second
order of the limits in Eq.~(\ref{WTBambiguity}). The only discussion or
motivation we can find is the statement reported in~(\ref{stateWTB}) which
suggests the (wrong) assumption of uniform convergence, so that the BK
teleportation channel $\mathcal{I}^{\mu}$ would \textquotedblleft
induce\textquotedblright\ the identity channel $\mathcal{I}$
(\textquotedblleft perfect quantum channel\textquotedblright) when
$\mu\rightarrow\infty$ (i.e., for infinite energy). We know that this is not
true, because%
\begin{equation}
\lim_{\mu\rightarrow\infty}\left\Vert \mathcal{I}^{\mu}-\mathcal{I}\right\Vert
_{\diamond}=2~.
\end{equation}

Let us stress that, in an infinite-dimensional Hilbert space,
considering ``an arbitrary protocol'' \cite{WildeFollowup} does
not necessarily mean that the protocol is energy-bounded. Among
the key-generation protocols to be considered in the derivation of
upper bounds for the (unconstrained) secret key capacity of
bosonic channels, one clearly needs to include protocols based on
asymptotic input states. The energy of the input state can diverge
in each single use of the protocol or as a monotonic function in
the number $n$ of the uses. For instance, one may just use TMSV
states with increasing energy as $\tilde{\mu}\simeq O(n)$ or any
other scaling in $n$. For any simulation energy $\mu$, we can
always consider such a diverging sequence, so that
$\varepsilon_{\text{TP}}(n,\mu)\overset{n}{\rightarrow}1$. In
other words, we generally have
\begin{equation}
\underset{n,\mu}{\lim\sup}~\varepsilon_{\text{TP}}(n,\mu)=1~.
\end{equation}
As a consequence, the joint limit for large $\mu$ and $n$ in the term
$\Delta(n,\mu)$ of Eq.~(\ref{followPLOB}) is not defined, and we can easily
get the explosion $\Delta(n,\mu)\rightarrow\infty$.

\subsection{Fixing the mathematical issues}

Let us know rigorously prove the WTB\ claim. We modify the derivation under
the assumption of an energy constraint on the input alphabet, which leads us
to write an energy-constrained version of Eq.~(\ref{followPLOB}). Only at the
very end, the enforced constraint can be relaxed once we have proven that the
upper bound does not depend on it. The key step is to prove a rigorous version
of Eq.~(\ref{pb2}) that removes the issue associated with the asymptotic
states (unbounded alphabet). The present proof is based on the bounded-uniform
convergence of the BK\ protocol. See Ref.~\cite{PLB} for other proofs that are
based on other topologies of convergence (e.g., strong convergence).

Following PLOB, let us restrict Alice's and Bob's registers to a
finite-energy alphabet $\mathcal{D}_{N}$ as in
Eq.~(\ref{finiteALFA}) where $N$ is maximum mean number of
photons. We then consider the energy-constrained diamond distance
$\left\Vert \cdot\right\Vert _{\diamond N}$ between a Gaussian
channel $\mathcal{E}$ and its simulation $\mathcal{E}^{\mu}$ which
defines a simulation error $\delta(\mu,N)$ as in
Eq.~(\ref{errorAA}). For any fixed energy, we may now state that
the simulation is asymptotically perfect, i.e.,
$\delta(\mu,N)\overset{\mu}{\rightarrow}0$ as in
Eq.~(\ref{defBBNN}).

The next step is to propagate this error to the output state as done in PLOB
and explained in Sec.~\ref{subsecSTRE}. For any energy constraint $N$ (bounded
alphabet) and finite-energy simulation $\mu$ of the Gaussian channel, we may
bound the trace distance between the actual output $\rho_{\mathbf{ab}}^{n}%
$\ and the simulated output $\rho_{\mathbf{ab}}^{\mu,n}$ as in
Eq.~(\ref{tossll}). In other words, we may use our peeling argument and write%
\begin{equation}
\delta(n,\mu,N):=\left\Vert \rho_{\mathbf{ab}}^{n}-\rho_{\mathbf{ab}}^{\mu
,n}\right\Vert \leq n\delta(\mu,N).
\end{equation}
Using the Fuchs-van der Graaf relation in Eq.~(\ref{FuchsVan}), we may now
correctly write the infidelity as
\begin{equation}
\varepsilon_{\text{TP}}(n,\mu,N):=1-F\left(  \rho_{\mathbf{ab}}^{n}%
,\rho_{\mathbf{ab}}^{\mu,n}\right)  \leq\frac{n\delta(\mu,N)}{2}~.
\end{equation}

Using the triangle inequality for the trace distance $d(\rho,\sigma
)=\sqrt{1-F(\rho,\sigma)}$, one finds that Eq.~(\ref{errorTT}) has to be
changed into the following%
\begin{equation}
\varepsilon(n,\mu,N):=\min\left\{  1,\left[  \sqrt{\varepsilon}+\sqrt
{\varepsilon_{\text{TP}}(n,\mu,N)}\right]  ^{2}\right\}  .
\end{equation}
As a result, for any $n$ and $N$, we may derive the energy-constrained version
of Eq.~(\ref{followPLOB}) which reads
\begin{equation}
K(\mathcal{E}|N)\leq\Phi(\mathcal{E})+\Delta(n,\mu,N),
\end{equation}
where $K(\mathcal{E}|N)$ is the reduced key rate associated with the use of an
energy-constrained alphabet, $\Phi(\mathcal{E})$ is the weak converse in PLOB,
and $\Delta$ has the asymptotic expansion
\begin{align}
\Delta(n,\mu,N)  &  \simeq\sqrt{\frac{2V(\mathcal{E})+O(\mu^{-1}%
)}{n[1-\varepsilon(n,\mu,N)]}}\nonumber\\
&  +\frac{C[\varepsilon(n,\mu,N)]}{n}+O(\mu^{-1})~,
\end{align}
with a more simplified expression for the pure-loss channel and the
quantum-limited amplifier.

Now, for any number of channel uses $n$ and any energy constraint $N$ for the
input alphabet, we may safely take the limit in $\mu$. In these conditions,
$\delta(\mu,N)\overset{\mu}{\rightarrow}0$ implies
\begin{equation}
\underset{\mu}{\lim\sup}~\varepsilon(n,\mu,N)=\varepsilon~,
\end{equation}
so that we may write the upper bound%
\begin{equation}
K(\mathcal{E}|N)\leq\Phi(\mathcal{E})+\Delta,~~\Delta=\sqrt{\frac
{2V(\mathcal{E})}{n(1-\varepsilon)}}+\frac{C(\varepsilon)}{n}~.
\end{equation}
Here we note that the bound $\Phi(\mathcal{E})+\Delta$ does not depend on $N$.
We can therefore relax the energy constraint by extending the inequality to
the supremum%
\begin{align}
K(\mathcal{E})  &  :=\sup_{N}K(\mathcal{E}|N)\nonumber\\
&  \leq\Phi(\mathcal{E})+\sqrt{\frac{2V(\mathcal{E})}{n(1-\varepsilon)}}%
+\frac{C(\varepsilon)}{n}~,
\end{align}
which rigorously proves the WTB claim (strong converse) for noisy Gaussian
channels. It is easy to see that we may similarly show the claim in
Eq.~(\ref{cc11}) for the pure-loss channel and the quantum-limited amplifier.

\subsection{Further details and technical issues}

There are other technical issues in WTB's treatment of bosonic Gaussian
channels that can be automatically fixed by correctly applying the tools in
PLOB. One of these issues is related with the treatment of the Bell detection
which is energy-unbounded for CV systems (being a projection onto
infinitely-squeezed displaced TMSV states). To be completely rigorous, this
measurement needs to be treated as a sequence of Gaussian measurements with
increasing energy. These measurements are quasi-projections onto
finite-squeezed displaced TMSV states $D(\alpha)\Phi^{\mu}D(-\alpha)$, with
$D(\alpha)$ being the displacement operator with amplitude $\alpha$~\cite{RMP}.

Thus, more precisely, the asymptotic simulation of a bosonic Gaussian channel
$\mathcal{E}$ must also involve a sequence of LOCCs $\mathcal{T}_{\mu}$
(including the finite-squeezing Bell measurements) which means that the
simulating channel $\mathcal{E}^{\mu}$ should be modified into the following
form~\cite{PLOB}
\begin{equation}
\mathcal{E}^{\mu}(\rho)=\mathcal{T}_{\mu}(\rho\otimes\rho_{\mathcal{E}}^{\mu
}),
\end{equation}
where $\rho_{\mathcal{E}}^{\mu}$\ is the usual quasi-Choi matrix. As a
consequence, the teleportation stretching of a protocol over a
bosonic\ Gaussian channel $\mathcal{E}$ involves a sequence of LOCCs
$\bar{\Lambda}_{\mu}$, so that it takes the form~\cite{PLOB}
\begin{equation}
||\rho_{\mathbf{ab}}^{n}-\bar{\Lambda}_{\mu}(\rho_{\mathcal{E}}^{\mu\otimes
n})||\overset{\mu}{\rightarrow}0,
\end{equation}
where we need to consider the simultaneous infinite-energy limit in both the
Choi-sequence $\rho_{\mathcal{E}}^{\mu}$ and the LOCCs $\bar{\Lambda}_{\mu}$.
For the sake of simplicity, we have omitted this further technical detail in
this review, but this aspect has been fully accounted in PLOB. Unfortunately,
no approximation of the CV Bell detection has been discussed or considered in
WTB. Furthermore, this problem also affects all the derivations presented in
Ref.~\cite{WildeComm}.

\section{Conclusion}

In this manuscript we have reviewed recent results and techniques in the field
of quantum and private communications. We have started with the definition of
adaptive protocols which are based on LOs assisted by two-way CCs. Optimizing
over these adaptive LOCCs, one can define the various two-way assisted
capacities ($Q_{2}$, $D_{2}$, $P_{2}$ and $K$) associated with a quantum
channel. In particular, the secret key capacity of the channel is defined
starting from the notion of private state which is a suitable cryptographic
generalization of a maximally-entangled state. Following PLOB, we have then
introduced the relative entropy of entanglement as a general weak converse
upper bound for the secret key capacity for quantum channels of any dimension
(finite or infinite). The first rigorous proof of this result was presented
back in 2015~\cite{PLOBv2} and exploits a truncation argument for the case of
CV channels. In this regard, we have also demystified some unfounded claims
made in recent literature.

We have then presented the most general kind of simulation for a quantum
channel in a quantum or private communication scenario. This must be based on
LOs performed by the two remote users. In fact, it is generally defined as an
LOCC applied to a resource state, and this formulation may also be asymptotic,
i.e., involving the limit of sequence of states, which is particularly
relevant for CV channels. Such a general LOCC simulation, first considered in
PLOB~\cite{PLOB}, has a number of precursory ideas based on teleportation that
have been developed in the last 20 years or
so~\cite{B2main,Niset,Leung,HoroTEL,Cirac,Wolfnotes,MHthesis,SougatoBowen}.

In the context of channel simulation, we have discussed the important
criterion of teleportation covariance, which is a way to determine if a
quantum channel is Choi-stretchable, i.e., simulable by teleportation over its
Choi matrix. This criterion was first identified for DV
channels~\cite{Wolfnotes,MHthesis,Leung} and then extended to channels of any
dimension~\cite{PLOB}. Most importantly, we have fully clarified how to handle
the asymptotic (and optimal) simulation of bosonic Gaussian channels, for
which the simulation error must be carefully controlled and correctly defined
in terms of energy-constrained diamond distance.

The tool of channel simulation is at the core of the most powerful techniques
of protocol reduction. This was first shown in the teleportation-based
approach of BDSW~\cite{B2main} with the formulation of a protocol reduction
into entanglement distillation that was later picked up by several other
works~\cite{Niset,Leung,MHthesis}. More recently, PLOB showed how channel
simulation (standard or asymptotic) is even more powerful and can be used to
reduce any adaptive protocol into a block protocol, while preserving the
original quantum task. This method of teleportation stretching has been
already widely exploited in recent literature, not only in the area of
quantum/private communication, but also in those of quantum channel
discrimination and quantum metrology (e.g., see Refs.~\cite{Metro,MetroREVIEW}).

With all the ingredients in our hands, we have discussed how their combination
leads to the computation of single-letter upper bounds for the two-way
capacities of quantum channels at any dimension (finite or infinite). Some of
these upper bounds coincide with corresponding lower bounds, and fully
establish the two-way capacities of fundamental quantum channels, such as the
lossy channel. In order to fully clarify the procedure, we have separately
discussed the results involving standard non-asymptotic simulations from those
that require asymptotic simulations (important for bosonic Gaussian channels).
While this recipe was designed in PLOB~\cite{PLOB} for the relative entropy of
entanglement, here we also discuss its full generality and applicability to
other entanglement measures, including the squashed entanglement.

There are a number of questions still open. What are the two-way capacities
($Q_{2}=D_{2}$ and $P_{2}=K$) of the depolarizing channel? Same question for
the amplitude damping channel. In the CV\ setting, the two-way capacities are
still to be determined for all the \textquotedblleft noisy\textquotedblright%
\ single-mode phase insensitive Gaussian channels, where the environment is
not just the vacuum. The most notable case is the thermal-loss channel for its
importance in QKD. From this point of view, this paper has also faced another
crucial question, what is the maximum excess noise that is tolerable in QKD?
Our study shows that the gap between upper and lower bound is still too large.
Perhaps these questions may be closed by following the approach recently put
forward in Ref.~\cite{PBTpaper}\ where port-based
teleportation~\cite{PBT,PBT1,PBT2,telereview} is adopted as more general tool
for channel simulation and protocol stretching.

In conclusion, we have also re-considered the derivations of the
follow-up work WTB~\cite{WildeFollowup}, which aimed at proving
the strong converse property of the previous upper bounds
established in PLOB. Because of a problem associated with the
unboundedness of the alphabet in the teleportation of CV channels,
the treatment of bosonic Gaussian channels was affected by a
technical issue that we fix in this paper. Furthermore, we also
fill a fundamental gap in the proof of WTB which was not properly
designed for adaptive protocols, due to the absence of a crucial
peeling argument~\cite{PLOB}. In this way, we provide a complete
and rigorous proof of the claims presented in WTB in relation to
the strong converse bounds for private communication over Gaussian
channels. Further validations can also be found in
Ref.~\cite{PLB}.

Let us conclude by saying that, despite the lack of technical rigor in
treating the simulation of bosonic Gaussian channels, we think that it is fair
to attribute to WTB~\cite{WildeFollowup} the derivation of their strong
converse bounds. By contrast, let us stress that WTB\ \textit{did not} play
any role in the derivation of the previous weak converse bounds (and two-way
assisted capacities) for the same channels, because these results were already
and rigorously established in PLOB~\cite{PLOB}, which also laid down the main methodology.

\section*{Acknowledgments}

This work has been supported by the Innovation Fund Denmark (Qubiz project),
the European Union (MSCA-IF-2016), and the EPSRC via the EPSRC grant
EP/K034480/1 and the `UK Quantum Communications Hub' (EP/M013472/1).

\appendix

\section{Fidelity limits in the BK teleportation protocol\label{App1}}

To explicitly show Eq.~(\ref{opp}), recall that a TMSV\ state $\Phi
_{Aa}^{\tilde{\mu}}$ is a bipartite Gaussian state with zero mean value and CM
of the form%
\begin{equation}
\mathbf{V}=\mathbf{V}_{q}\oplus\mathbf{ZV}_{q}\mathbf{Z,} \label{formCM}%
\end{equation}
where $\mathbf{Z}:=\mathrm{diag}(1,-1)$ and $\mathbf{V}_{q}$ is
explicitly
given by%
\begin{equation}
\mathbf{V}_{q}(\tilde{\mu})=\left(
\begin{array}
[c]{cc}%
\tilde{\mu} & \sqrt{\tilde{\mu}^{2}-1/4}\\
\sqrt{\tilde{\mu}^{2}-1/4} & \tilde{\mu}%
\end{array}
\right)  ~.
\end{equation}

Now assume that we apply the BK protocol to mode $a$ of the input state
$\Phi_{Aa}^{\tilde{\mu}}$ by using a TMSV state $\Phi_{a^{\prime}B}^{\mu}$ as
a resource. The ideal CV\ Bell detection on modes $a$ and $a^{\prime}$, and
the CC of the outcome realizes the BK channel $\mathcal{I}^{\mu}$ from mode
$a$ to mode $B$. This is locally (i.e., point-wise) equivalent to an
additive-noise Gaussian channel with added noise~\cite{GerLimited,RicFINITE}
$\xi=2\mu-\sqrt{4\mu^{2}-1}$. When applied to $\Phi_{Aa}^{\tilde{\mu}}$, we
get the output $\Phi_{Aa}^{\mu,\tilde{\mu}}:=\mathcal{I}_{A}\otimes
\mathcal{I}_{a}^{\mu}(\Phi_{Aa}^{\tilde{\mu}})$ whose CM $\mathbf{V}%
^{\mu,\tilde{\mu}}$ has the form in Eq.~(\ref{formCM}) with
\begin{equation}
\mathbf{V}_{q}^{\mu,\tilde{\mu}}=\left(
\begin{array}
[c]{cc}%
\tilde{\mu} & \sqrt{\tilde{\mu}^{2}-1/4}\\
\sqrt{\tilde{\mu}^{2}-1/4} & \tilde{\mu}+\xi
\end{array}
\right)  ~.
\end{equation}

Using the formula for the quantum fidelity between arbitrary multimode
Gaussian states~\cite{banchiPRL2015}, we find
\begin{gather}
F(\mu,\tilde{\mu}):=F\left(  \Phi_{Aa}^{\mu,\tilde{\mu}},\Phi_{Aa}^{\tilde
{\mu}}\right) \\
=\frac{1}{\sqrt[4]{1-4\tilde{\mu}\left[  \sqrt{4\mu^{2}-1}+\tilde{\mu}%
-2\mu\left(  1+2\tilde{\mu}\xi\right)  \right]  }}.\nonumber
\end{gather}
We can easily check the asymptotic expansions%
\begin{align}
F(\mu,\tilde{\mu})  &  \simeq1-O(\mu^{-1}),~~\text{for large }\mu,\\
F(\mu,\tilde{\mu})  &  \simeq O(\tilde{\mu}^{-1}),~~\text{for large }%
\tilde{\mu},
\end{align}
which imply the opposite limits in Eqs.~(\ref{FidLIMIT}) and~(\ref{opp}), when
$\mathcal{E}=\mathcal{I}.$


\begin{thebibliography}{999}
\bibitem {Eps}More precisely, this defines an $(n,R_{n},\varepsilon
)$-protocol, but we omit this\ technical notation here in order to simplify
the physical discussion.
\bibitem {details}More precisely, suppose to run a key generation protocol
$\mathcal{P}_{n}$ for large number of uses $n$ and with rate $R_{n}$. For any
$\varepsilon>0$ there is an $n_{0}$ such that the truncated protocol
$\mathcal{P}_{n_{0}}$ has rate $R_{n_{0}}\geq R_{n}-\varepsilon$. Repeat the
truncated protocol $m=n/n_{0}$ times, so that Alice and Bob collect $m$ copies
of the state $\rho_{\mathbf{ab}}^{n_{0}}$. Performing one-way key distillation
over these copies, they achieve a key rate~\cite{Matthias1a,Matthias2a}
$\tilde{R}_{n}\geq(1-8\varepsilon)(R_{n}-\varepsilon)-4n_{0}^{-1}%
H_{2}(\varepsilon)$, where $H_{2}$ is the binary Shannon entropy. The overall
protocol $\mathcal{\tilde{P}}=\mathcal{P}_{n_{0}}^{\otimes m}$ has a rate
$\tilde{R}_{n}$ approaching $R_{n}$ by decreasing $\varepsilon$ (and
correspondingly increasing $n_{0}$). At the same time, its classical
communication cost is at most linear in $m$~\cite{DeveWinter}, which
implies~\cite{KD} that the shield size $d_{\mathrm{S}}$ increases at most
exponentially in $m<n$.
\bibitem {ClaimsW}In Ref.~\cite{WildeQIP}, one may read: \textquotedblleft It
is worthwhile to point out that no such issue [unbounded shield] exists in
various proofs that the relative entropy of entanglement is an upper bound on
distillable key [17,18,22,40]\textquotedblright\ where [22] is PLOB.
\bibitem {SimLOCC}Note that, more generally, one also needs to consider
sequences of LOCCs $\mathcal{T}^{\mu}$, so that the asymptotic simulation
reads $\mathcal{E}(\rho)=\lim_{\mu}\mathcal{T}^{\mu}(\rho\otimes\sigma^{\mu}%
)$. For simplicity we omit this technical generalization, referring the reader
to Ref.~\cite{PLOB} for more details. See also the discussion at the end of
this paper (Sec.~X E).
\end{thebibliography}
\end{document}